\documentclass[11pt]{article}
\usepackage[utf8]{inputenc}
\usepackage[letterpaper,margin=1in]{geometry}
\usepackage{authblk}

\usepackage[shortlabels]{enumitem}

\usepackage[numbers,sort]{natbib}

\usepackage{amsfonts,amssymb,amsmath,amsthm}
\usepackage{mathtools}
\usepackage{booktabs,tabularx}
\usepackage{microtype}

\usepackage[table,dvipsnames]{xcolor}
\usepackage{tikz}
\usetikzlibrary{arrows.meta}
\usepackage[colorlinks=true,pdfpagemode=UseNone,urlcolor=RoyalBlue,linkcolor=RoyalBlue,citecolor=OliveGreen,pdfstartview=FitH]{hyperref}
\usepackage{cleveref}

\usepackage{thmtools}
\usepackage{thm-restate}

\declaretheorem{theorem}
\declaretheorem[numberwithin=section]{lemma}
\declaretheorem[sibling=lemma]{proposition}
\declaretheorem[sibling=lemma]{corollary}
\declaretheorem[sibling=lemma,style=remark]{remark}
\AddToHook{env/theorem/begin}{\crefalias{section}{theorem}}
\AddToHook{env/lemma/begin}{\crefalias{section}{lemma}}
\AddToHook{env/proposition/begin}{\crefalias{section}{proposition}}
\AddToHook{env/corollary/begin}{\crefalias{section}{corollary}}
\AddToHook{env/remark/begin}{\crefalias{section}{remark}}

\hypersetup{pdftitle={The Optimizer's Windfall: Bounding Hindsight Gain under Random Restrictions}}
\crefname{section}{Section}{Sections}
\crefname{subsection}{Section}{Sections}
\crefname{theorem}{Theorem}{Theorems}
\crefname{lemma}{Lemma}{Lemmas}
\crefname{proposition}{Proposition}{Propositions}
\crefname{corollary}{Corollary}{Corollaries}
\crefname{remark}{Remark}{Remarks}
\crefname{enumi}{item}{items}
\crefname{appendix}{Appendix}{Appendices}
\crefname{table}{Table}{Tables}
\crefname{figure}{Figure}{Figures}
\newcommand{\E}{\mathbb{E}}
\newcommand{\Prob}{\mathbb{P}}
\newcommand{\R}{\mathbb{R}}
\newcommand{\eps}{\varepsilon}
\newcommand{\one}{\mathbf{1}}
\newcommand{\cA}{\mathcal{A}}
\newcommand{\cD}{\mathcal{D}}
\newcommand{\cE}{\mathcal{E}}
\newcommand{\cS}{\mathcal{S}}
\newcommand{\ip}[2]{\langle #1,#2\rangle}
\newcommand{\norm}[1]{\lVert #1\rVert}
\newcommand{\abs}[1]{\lvert #1\rvert}
\newcommand{\fro}[1]{\norm{#1}_{\mathrm{F}}}
\newcommand{\opn}[1]{\norm{#1}_{\mathrm{op}}}
\newcommand{\dTV}{d_{\mathrm{TV}}}
\DeclareMathOperator{\Val}{Val}
\DeclareMathOperator{\Opt}{Opt}
\DeclareMathOperator{\mc}{mc}
\DeclareMathOperator{\conv}{conv}
\DeclareMathOperator{\Var}{Var}
\DeclareMathOperator{\tr}{tr}

\DeclareMathOperator{\Gap}{Gap}

\title{The Optimizer's Windfall:\\[2pt] Bounding Hindsight Gain under Random Restrictions}
\author{Xiaohui Bei}
\author{Wenhao Wu}
\affil{
  Nanyang Technological University\\ Singapore\\
  \texttt{xhbei@ntu.edu.sg} \qquad \texttt{wenhao008@e.ntu.edu.sg}
}
\date{}

\begin{document}
\maketitle
\vspace{-20pt}
\begin{abstract}
Suppose an optimization problem is subjected to a random restriction on each variable independently, and is re-optimized after the restrictions are revealed. For multilinear objectives, every fixed solution is unbiased after appropriate rescaling, but choosing a solution after seeing the restriction introduces an upward bias. We bound this excess, called the \emph{hindsight gain}, over products of arbitrary compact local domains in Euclidean unit balls.

For interactions $A=(A_e)_e$ among $r$ distinct variables, and independent random weights $\zeta_i$ being applied to each variable $i$, we show that the gain is at most $O_r(\fro A M^{(r-1)/2}\sigma_{\max})$, where $\fro A^2$ is the sum of squares of all interaction coefficients, $M=\sum_i\E\zeta_i^2$, and $\sigma_{\max}=\max_i\sqrt{\Var(\zeta_i)}$. This bound requires only second moments. For uniform restriction, where each variable is retained with the same probability $p$, we further sharpen the bound to $O_r(p^{r-1/2}\sqrt{1-p}\,n^{(r-1)/2}\fro A)$, and show that this dependence is tight for every fixed $r$ and $p\in(0,1)$. Our proof applies a deterministic comparison principle that bounds the difference of two optimal values through a single support direction of their feasible sets. We then combine tensor projection with vector contraction to obtain the second-moment bound, and sharpen it for uniform restriction by repeated thinning and exact factorial-moment calculations.

These bounds lead to several applications. (i)~For every fixed arity $r$ and alphabet size $k$, we show that sampling $O(\eps^{-2})$ variables and solving the induced subproblem suffices to estimate a dense Max-$r$CSP optimum within $\pm\eps n^r$, and this sampling exponent is optimal. (ii)~The same approach estimates MaxCut and the distance to bipartiteness within $\pm\eps n^2$ using $O(\eps^{-4})$ nonadaptive queries and $O(\eps^{-2})$ examined vertices, both optimal in their respective models. (iii)~We answer a question of Yufei Zhao: vertex percolation at rate $p$ preserves the max-cut fraction whenever $p\,e(G)/v(G)\to\infty$, and this condition is sharp. (iv)~For bounded $r$-local quantum Hamiltonians with fixed local dimension $d$, $O(\eps^{-3})$ uniformly sampled particles suffice to estimate the ground-state energy within $\pm\eps n^r$, improving the previous $O(\eps^{-6}\log(1/\eps))$ bound.
\end{abstract}
\clearpage
\tableofcontents
\clearpage
\section{Introduction}\label{sec:intro}

A recurring paradigm in sublinear algorithms, property testing, and probabilistic combinatorics is to \emph{restrict} an optimization problem to a random subset of its variables, solve the restricted problem, and rescale. The prototypical example is MaxCut in dense graphs, for which sampling-based algorithms have been studied since the 1990s~\cite{AKK99,dlV96,FK99,GGR98}. Under this regime, the algorithm retains each vertex of an $n$-vertex graph $G$ independently with probability $p$, obtaining a random set $S$, computes the maximum cut $\mc(G[S])$ for the induced subgraph $G[S]$, and outputs $p^{-2}\mc(G[S])$. This natural estimator was analyzed in~\cite{AFKK03,MS08,RV07}. Note that every \emph{fixed} cut keeps each of its edges with probability exactly $p^2$, so its rescaled size is an unbiased estimate of its true size. The maximum cut of $G[S]$, however, is chosen \emph{after} $S$ is revealed. It need not be the restriction of any good cut of $G$, and is always at least as large as the restriction of a maximum cut of $G$. The estimator is therefore biased upward, and its bias is exactly the value of being allowed to re-optimize after seeing the restriction.
The triangle already exhibits the effect. For $G=K_3$,
\[
\E\mc(K_3[S]) = 3p^2(1-p)\cdot 1+p^3\cdot 2 = 3p^2-p^3,
\qquad\text{whereas}\qquad
p^2\mc(K_3)=2p^2 .
\]
The excess $p^2(1-p)$ comes from the event that exactly two vertices survive: the surviving edge is then cut, although no cut of $K_3$ cuts all three edges.

More generally, let $f_\zeta$ be a random objective with pointwise mean $f(x)=\E f_\zeta(x)$. We call the nonnegative quantity
\[
\E\Bigl[\max_x f_\zeta(x)\Bigr]-\max_x f(x)
\]
the \emph{hindsight gain}. This notion appears in many contexts. In decision analysis, it is the selection bias of the maximum of a random field~\cite{SW06}. In information theory it is the Bayesian regret of a greedy policy under the mean objective~\cite{RZ16}.
In stochastic programming it is known as the expected value of perfect information: the difference between the ``wait-and-see'' and the ``here-and-now'' optimum~\cite{AW70,BL11}. It is also the optimistic bias of sample average approximation, which decreases as the sample grows~\cite{MMW99}. For binary quadratic programs with random coefficients, the wait-and-see value has been bounded using only the marginal distributions of the coefficients~\cite{NST18}. In sampling-based estimation it is the bias of the sample-and-solve estimator, and bounding it is the main step in showing that a random restriction preserves an optimum.

In this paper we bound the hindsight gain for homogeneous multilinear optimization over arbitrary compact local domains, for every fixed degree and uniform retention rate $p\in(0,1)$, and derive several algorithmic and combinatorial consequences.

\subsection{The Model: Multilinear Interactions}\label{sec:model}

We consider an optimization problem with $n$ variables, where variable $i$ chooses a vector $x_i$ from a nonempty compact subset $K_i$ of the Euclidean unit ball in $\R^D$. The set $K_i$ specifies the allowed states; it need not be convex or equal to the whole unit ball. Write $K:=\prod_iK_i$. The dimension $D$ may depend on the instance, and variables with different dimensions can be accommodated by padding their vectors with zeros.

Fix an integer $r\ge2$. For each $r$-element set $e=\{i_1<\cdots<i_r\}\subseteq[n]$, let $A_e\colon(\R^D)^r\to\R$ be an interaction that is linear in each argument separately. Missing interactions are considered to be zero. With $x_e=(x_{i_1},\ldots,x_{i_r})$, we write the objective and its optimum as
\begin{equation}\label{eq:local-objective}
f_A(x):=\sum_{e\in\binom{[n]}r}A_e(x_e),
\qquad
\Val_A:=\max_{x\in K}f_A(x).
\end{equation}
Thus every interaction involves exactly $r$ distinct variables. For example, when $r=2$, each interaction is given by a matrix $A_{ij}$ and
$f_A(x)=\sum_{i<j}x_i^\top A_{ij}x_j$.
For general $r$, the coefficients of $A_e$ form an $r$-way array.

Next, we consider a random noise being applied to the variables. Let $\boldsymbol{\zeta}=(\zeta_1,\ldots,\zeta_n)$, where each $\zeta_i$ is an independent nonnegative random weight with finite second moments. These weights multiply interaction $e$ by the product of its variable weights. The weighted objective and its optimum become

\begin{equation}\label{eq:weighted-objective}
f_{A,\boldsymbol{\zeta}}(x):= \sum_{e\in\binom{[n]}r}\Bigl(\prod_{i\in e}\zeta_i\Bigr)A_e(x_e),
\qquad 
\Val_A(\boldsymbol{\zeta}):=\max_{x\in K}f_{A,\boldsymbol{\zeta}}(x).
\end{equation}

We use the same notation for deterministic nonnegative weights; in particular, $\Val_A(\one)=\Val_A$.
If we further put $p_i=\E\zeta_i$ and $\mathbf{p}=(p_i)_i$, we can define the \emph{hindsight gain} by
\begin{equation}\label{eq:gap}
\Gap_A(\boldsymbol{\zeta}):=\E\Val_A(\boldsymbol{\zeta})-\Val_A(\mathbf{p}).
\end{equation}

For every fixed assignment $x\in K$, independence gives $\E f_{A,\boldsymbol{\zeta}}(x)=f_{A,\mathbf{p}}(x)$. Thus the hindsight gain is nonnegative: it measures the expected benefit of choosing an assignment after observing the weights, compared with optimizing the mean objective.

This model includes several important special cases. It covers every Max-$r$CSP over finite alphabets: $K_i$ consists of the coordinate unit vectors representing its labels, and $A_e$ has the payoff table as its coefficients (\cref{sec:csp}). MaxCut is a two-label example with $r=2$. The model also includes continuous vector-spin objectives and the negative of the minimum product-state energy of an $r$-local quantum Hamiltonian, using real coordinate vectors of single-particle density matrices (\cref{sec:quantum}).

\paragraph{Why multilinear interactions.}
We require every interaction to be multilinear in its involved variables so that independent weights give $\E f_{A,\boldsymbol{\zeta}}(x)=f_{A,\mathbf{p}}(x)$ for every fixed assignment $x$, and hence a nonnegative hindsight gain. This assumption also makes the objective affine in each variable separately, allowing us to replace each local domain by its convex hull without changing any weighted or restricted optimum; we rely on these two properties in proving the general bound in \cref{thm:noise} and \cref{thm:uniform}. 

\subsection{Results: Bounds on the Hindsight Gain}\label{sec:results}

Our main results in this paper bound the hindsight gain for multilinear objectives of fixed degree. We first give a general bound that applies to arbitrary independent nonnegative weights.

\begin{restatable}[Independent weights]{theorem}{ThmNoise}\label{thm:noise}
Let $A=(A_e)_e$ be a collection of $r$-variable multilinear interactions in the model of \cref{sec:model}, and let $\zeta_1,\ldots,\zeta_n$ be independent nonnegative random variables with finite second moments. Then
\[
0\le\Gap_A(\boldsymbol{\zeta})
\le r\sqrt5\,\fro A\,M^{(r-1)/2}\sigma_{\max},
\]
where
$
\fro A^2:=\sum_{e\in\binom{[n]}r}\fro{A_e}^2,\quad M:=\sum_i\E\zeta_i^2,\quad \sigma_{\max}:=\max_i\sqrt{\Var(\zeta_i)}.
$
Here $\fro{A_e}$ is the Frobenius norm of the coefficient array of $A_e$.
\end{restatable}

The bound depends only on the coefficient norm, the degree, and two moment parameters. It is independent of the vector dimension and of the cardinalities or covering numbers of the local feasible sets. In particular, interactions of any fixed degree require only second moments of the weights. This theorem also covers importance sampling, where a retained variable is reweighted by the reciprocal of its inclusion probability. A finer estimate (\cref{thm:refined}) weights each variance by the sum of squared coefficients of the interactions involving that variable.

\medskip

As mentioned at the beginning, a common paradigm is to restrict an optimization problem to a random subset of its variables, solve the restricted problem, and rescale. This corresponds to applying \cref{thm:noise} to Bernoulli weights $\zeta_i\sim\mathrm{Bernoulli}(p_i)$, which retain each variable $i$ independently with probability $p_i$.
We write $\Val_A(S):=\Val_A(\one_S)$ for the optimum restricted to $S$. Then \cref{thm:noise} gives
\begin{equation}\label{eq:bern-intro}
0\le\E\Val_A(S)-\Val_A(\mathbf{p})
\le r\sqrt5\,\fro A\left(\sum_i p_i\right)^{(r-1)/2}
\sqrt{\max_i p_i(1-p_i)}.
\end{equation}

For uniform retention, take $p_i=p$ for every $i$, so $\mathbf{p}=p\one$, and write $S_p$ for the resulting random set. Homogeneity gives $\Val_A(p\one)=p^r\Val_A$, so
\[
\Gap_A(\one_{S_p})=\E\Val_A(S_p)-p^r\Val_A.
\]
In this case, \cref{thm:noise} gives
$\Gap_A(\one_{S_p})\le r\sqrt5\,p^{r/2}\sqrt{1-p}\,n^{(r-1)/2}\fro A$.
Let $q=pn$ denote the expected sample size. At the coefficient scale $\fro A=O_r(n^{r/2})$, the resulting normalized bias satisfies
\[
p^{-r}\Gap_A(\one_{S_p})
=O_r\!\left(n^r\frac{n^{(r-1)/2}}{q^{r/2}}\right).
\]
Relative to the natural objective scale $n^r$, this bound still grows with $n$ when $q$ is fixed. In our next theorem, we remove this dependence and prove a sharper bound of $O_r(q^{-1/2})$ on the relative bias.

\begin{restatable}[Uniform retention]{theorem}{ThmUniform}\label{thm:uniform}
For every $r\ge2$ there is a constant $C_r$ such that, for every collection $A$ in the model of \cref{sec:model} and every $p\in(0,1]$,
\[
0\le\E\Val_A(S_p)-p^r\Val_A
\le C_r p^{r-1/2}\sqrt{1-p}\,n^{(r-1)/2}\fro A.
\]
One may take $C_r=8r\,2^{r/2}$.
\end{restatable}

Equivalently, the normalized bias is at most
$C_r\sqrt{(1-p)/p}\,n^{(r-1)/2}\fro A$.
When $\fro A=O_r(n^{r/2})$, this is $O_r(n^r/\sqrt q)$: the bound on the bias relative to the natural objective scale $n^r$ depends only on the expected sample size $q$. 
This allows us to sample a small random subset of variables and still estimate the optimum of a dense multilinear objective within a small additive error, as we illustrate in \cref{sec:apps}.

\subsection{Applications}\label{sec:apps}

In this section we describe several applications of the hindsight-gain bounds. We first summarize the consequences in \cref{tab:applications}, and then give more details for each application.

\begin{table}[htbp]
\centering
\small
\renewcommand{\arraystretch}{1.25}
\begin{tabularx}{\textwidth}{>{\raggedright\arraybackslash}p{0.27\textwidth} >{\raggedright\arraybackslash}X >{\raggedright\arraybackslash}p{0.25\textwidth}}
\toprule
Application & Previous bounds & This work \\
\midrule
\rowcolor{gray!10}
Max-$r$CSPs
& $O(\eps^{-4})$~\cite{MS08}
& $\Theta(\eps^{-2})$ \\
MaxCut and bipartiteness\newline (vertex samples)
& $O(\eps^{-4})$~\cite{MS08,RV07}
& $\Theta(\eps^{-2})$ \\
MaxCut and bipartiteness\newline (nonadaptive edge queries)
& $\tilde O(\eps^{-6})$~\cite{AFKK03,GMRS22,MS08}
& $\Theta(\eps^{-4})$ \\
\rowcolor{gray!10}
Max-cut fraction under vertex percolation
& $e(G)\ge v(G)^{1.8}$,\newline $p=1/2$~\cite{GS25}
& $p\,e(G)/v(G)\to\infty$\newline (sharp) \\
Product-state energy
& $O(\eps^{-6}\log(1/\eps))$~\cite{Bergamaschi23}
& $\Theta(\eps^{-2})$ \\
\rowcolor{gray!10}
Quantum ground-state energy
& $O(\eps^{-6}\log(1/\eps))$~\cite{Bergamaschi23}
& $O(\eps^{-3})$ \\
\bottomrule
\end{tabularx}
\caption{Sample and edge-query bounds for additive error $\eps n^r$ ($\eps n^2$ for graph estimation), with constant success probability, fixed arity $r$, constant-size alphabets, and a fixed single-particle state space. 
The percolation row instead compares sufficient conditions for preserving the max-cut fraction.}
\label{tab:applications}
\end{table}

\paragraph{Sampling dense Max-$r$CSPs.}
Our first application is to determine how many variables we need to sample to estimate the optimum of a dense constraint satisfaction problem. A Max-$r$CSP has variables $1,\ldots,n$, a finite alphabet $\cA_i$ for each variable $i$, and one payoff table $g_e\colon\prod_{i\in e}\cA_i\to\R$ for each $r$-element set $e\subseteq[n]$. For $S\subseteq[n]$, we let $\Opt_g(S)$ be the maximum of $\sum_{e\subseteq S}g_e(x_e)$ over assignments on $S$, and write $\Opt_g=\Opt_g([n])$. For payoffs in $[0,1]$, our goal is to estimate $\Opt_g$ within $\pm\eps n^r$ from the instance induced by a small random subset of variables.

For this sampling problem, Alon, Fernandez de la Vega, Kannan, and Karpinski~\cite{AFKK03} proved an $O_r(\eps^{-4}\log(1/\eps))$ bound for Boolean Max-$r$CSPs. Subsequently, Mathieu and Schudy~\cite{MS08} obtained $O(\eps^{-4})$ for MaxCut and state that their result extends to dense Max-$r$CSPs; Rudelson and Vershynin~\cite{RV07} obtained the same exponent for Boolean Max-2CSPs.

For fixed arity and constant-size alphabets, we reduce the sample size to $O(\eps^{-2})$ by combining \cref{thm:uniform} with a sampling-sensitive variance bound, improving the previous bounds. 
Furthermore, we show that this $O(\eps^{-2})$ sample bound is tight in its dependence on $\eps$ for every $r\ge2$, even for Boolean alphabets. We state the sampling guarantee in the following theorem; the matching lower bound is proved in \cref{cor:csp-vertices}.

\begin{theorem}[Sampling dense Max-$r$CSPs]\label{thm:csp-intro}
For every fixed $r\ge2$ and fixed bound on the alphabet sizes, there is a constant $C$ such that the following holds. Let $g$ be a Max-$r$CSP on $n$ variables with alphabets satisfying this bound and payoffs in $[0,1]$, and let $\eps\in(0,1)$. If $S$ retains each variable independently with probability $p=\min\{1,C\eps^{-2}/n\}$, then, with probability at least $2/3$,
\[
\abs S\le4C\eps^{-2},
\qquad
\abs{p^{-r}\Opt_g(S)-\Opt_g}\le\eps n^r.
\]
\end{theorem}

To apply our framework, we use one-hot encoding to turn each payoff table into a multilinear interaction. After subtracting the mean of each table, the relevant coefficient norm is the \emph{centered energy} $\cE(g)$, defined as the Euclidean norm of all centered table entries (\cref{sec:encoding}). Our energy-sensitive estimate (\cref{thm:csp-estimator}) also records the actual table norms, which can substantially reduce the alphabet dependence.

\paragraph{MaxCut and the distance to bipartiteness.}
We next consider estimating the maximum cut of a graph and its distance to bipartiteness. We write $e(G)$ and $v(G)$ for the numbers of edges and vertices of $G$, and $\beta(G):=e(G)-\mc(G)$ for the distance to bipartiteness, that is, the minimum number of edges whose deletion makes $G$ bipartite. Our goal is to estimate $\mc(G)$ or $\beta(G)$ within $\pm\eps n^2$ for an $n$-vertex graph $G$.

To measure the cost of estimation, we work in the adjacency-matrix model, where an algorithm may ask whether a pair $\{u,v\}$ of vertices is an edge. It \emph{examines} a vertex if the vertex appears in one of its edge queries, and it is \emph{nonadaptive} if it fixes all its queries before seeing any answer. Reading the subgraph induced by $q$ sampled vertices therefore costs $\binom q2$ nonadaptive edge queries.

We first compare the number of sampled vertices. Rudelson and Vershynin~\cite{RV07} and Mathieu and Schudy~\cite{MS08} obtained an $O(\eps^{-4})$ vertex-sampling bound for MaxCut. By also estimating the edge count, we obtain the same sampling bound for the distance to bipartiteness.

If we instead count individual edge queries, the best previous bound that we are aware of is $\tilde O(\eps^{-6})$, obtained from the sampling results of~\cite{AFKK03,MS08} as explained in~\cite[Appendix~A]{GMRS22}; earlier, Goldreich, Goldwasser, and Ron~\cite{GGR98} estimated $\mc(G)$ with $\tilde O(\eps^{-7})$ edge queries.\footnote{A preprint~\cite{GMRS21} claimed an estimator for $\mc(G)$ that examines $\tilde O(\eps^{-3})$ vertices and makes $\tilde O(\eps^{-4})$ queries. It has since been withdrawn; according to the withdrawal notice, the proof of the main claim is incomplete.} For the related but weaker testing task, Ghosh, Mishra, Raychaudhury, and Sen~\cite{GMRS22} obtained $\tilde O(\eps^{-3})$ queries for the \emph{tolerant testing} problem of distinguishing $\beta(G)\le\eps n^2$ from $\beta(G)\ge(2+\Omega(1))\eps n^2$.

By applying \cref{thm:uniform} to the adjacency-matrix model, we obtain a nonadaptive algorithm that estimates both quantities within $\pm\eps n^2$, examining only $O(\eps^{-2})$ vertices and making $O(\eps^{-4})$ edge queries.
Moreover, we prove that both bounds are optimal: the nonadaptive query complexity is $\Theta(\eps^{-4})$, and the number of examined vertices is $\Theta(\eps^{-2})$, even when adaptive queries are allowed. We summarize these guarantees in the following theorem.

\begin{theorem}[MaxCut and bipartiteness]\label{thm:bip-intro}
Let $\eps\in(0,1)$.
\begin{enumerate}[(a)]
\item\label{it:bip-upper} There is a nonadaptive algorithm that, given an $n$-vertex graph $G$, examines $O(\eps^{-2})$ randomly chosen vertices and queries only pairs among them. With probability at least $2/3$, it outputs estimates of $\mc(G)$ and of $\beta(G)$ that are both accurate within $\pm\eps n^2$. It makes $O(\eps^{-4})$ edge queries.
\item\label{it:bip-lower} There are absolute constants $c,C>0$ such that the following holds for $\eps\le1/64$ and $n\ge C\eps^{-2}$. Every nonadaptive algorithm that estimates $\mc(G)$, or $\beta(G)$, within $\pm\eps n^2$ with probability at least $2/3$ on every $n$-vertex graph makes at least $c\,\eps^{-4}$ queries. Every algorithm with the same guarantee, adaptive or not, examines at least $c\,\eps^{-2}$ vertices in the worst case.
\end{enumerate}
\end{theorem}

For adaptive algorithms, our lower bound on examined vertices also implies that $\Omega(\eps^{-2})$ queries are necessary. 

\paragraph{Max-cut under vertex percolation.}
We next ask when vertex sampling preserves the fraction of edges in a maximum cut.
At the 2019 Oberwolfach workshop \emph{Combinatorics, Probability and Computing}~\cite{OWR19}, Yufei Zhao asked the following question: when does one-half vertex percolation of a graph $G$, which keeps each vertex independently with probability $\frac12$, preserve the max-cut fraction $\mc(G)/e(G)$ of graph $G$? Zhao observed that the answer is affirmative when $G$ is sufficiently dense, for example when $e(G)\ge v(G)^{2-c}$ for some constant $c>0$, and he asked whether the same holds for every sequence of graphs $G_N$ with $e(G_N)/v(G_N)\to\infty$.
For $p=1/2$, the previous sufficient condition $e(G)\ge v(G)^{1.8}$ (that is, $c=0.2$ in the exponent $2-c$) is recorded in the survey of Gishboliner and Shapira~\cite[p.~10]{GS25}. 

By applying \cref{thm:uniform} to MaxCut, we are able to answer this question completely in the affirmative, even for arbitrary retention probabilities.

\begin{theorem}[Max-cut under vertex percolation]\label{thm:zhao-intro}
Let $(G_N)$ be graphs with $e(G_N)\ge 1$, let $\rho_N:=e(G_N)/v(G_N)$, and let $p_N\in(0,1]$. If $p_N\rho_N\to\infty$, then with probability tending to one the random induced subgraph $H_N:=G_N[S_{p_N}]$ has at least one edge, and
\[
\frac{\mc(H_N)}{e(H_N)}-\frac{\mc(G_N)}{e(G_N)}\longrightarrow0\qquad\text{in probability.}
\]
Conversely, for every constant $c>0$ there are graphs $G_N$ with $\rho_N\to\infty$ and probabilities $p_N$ with $p_N\rho_N=c$ for which this conclusion fails.
\end{theorem}

Taking $p_N=\frac12$ answers Zhao's question. 
The same argument also shows that for every graph of average degree $\bar d$, uniform vertex sampling at rate $p=\min\{1,C\eps^{-2}/\bar d\}$ estimates the max-cut within a factor $1\pm\eps$ after rescaling by $p^{-2}$, with constant probability; this complements the core-sets of Bhaskara, Daruki, and Venkatasubramanian~\cite{BDV18} (\cref{rem:coreset}).

\paragraph{Sampling quantum ground-state energies.}
In our final application, we ask how many particles we need to sample to estimate the ground-state energy of a quantum Hamiltonian. This application uses the continuous local domains allowed by our model.

The ground-state energy of a quantum Hamiltonian is its minimum eigenvalue, the lowest energy of the corresponding many-particle system. Consider an $r$-local Hamiltonian
\[
H=\sum_{\substack{e\subseteq[n]\\1\le\abs e\le r}}h_e
\]
on $n$ particles with a fixed single-particle state space, where each Hermitian term $h_e$ acts on the particles in $e$, and $\opn{h_e}\le1$. There is at most one term per subset, and missing interactions are represented by zero terms. We write $E_0(H):=\lambda_{\min}(H)$ and $H_S:=\sum_{e\subseteq S}h_e$. Our goal is to estimate $E_0(H)$ within $\pm\eps n^r$ from the sampled Hamiltonian $H_S$, allowing arbitrary entanglement. We also consider the minimum product-state energy, where we restrict the state to a tensor product of single-particle states.

For this sampling problem, Bergamaschi~\cite[Theorem~11]{Bergamaschi23} proved that $O(\eps^{-6}\log(1/\eps))$ sampled particles suffice. The same sample bound for the minimum product-state energy is implicit in Sections~7 and~C.3 of his full version.

We improve these bounds to $O(\eps^{-2})$ samples for the minimum product-state energy and $O(\eps^{-3})$ samples for the full ground-state energy. To do so, we first apply \cref{thm:uniform} to the continuous space of single-particle density matrices. We then use Bergamaschi's product-state approximation theorem~\cite[Theorem~3]{Bergamaschi23}, building on Brand\~ao and Harrow~\cite{BH13}, to transfer the estimate to the full ground-state energy. We summarize the resulting guarantee in the following theorem.

\begin{restatable}[Sampling quantum ground-state energies]{theorem}{ThmQuantum}\label{thm:quantum-intro}
For every fixed integer $r\ge2$ and fixed single-particle state space, there is a constant $C$ such that the following holds. Let $H$ be an $r$-local Hamiltonian as above, and let $\eps\in(0,1)$. Put $q:=\min\{n,\lceil C\eps^{-3}\rceil\}$ and choose $S\subseteq[n]$ uniformly among its $q$-element subsets. Then, with probability at least $0.99$,
\[
\left|\left(\frac nq\right)^rE_0(H_S)-E_0(H)\right|\le\eps n^r.
\]
\end{restatable}

In obtaining this guarantee, we absorb the $O_r(n^r/q)$ sampling error from terms involving fewer than $r$ particles into the bound. Our estimator uses $O(\eps^{-3r})$ nonadaptive queries, each returning the classical description of a local interaction; its computation on the sampled Hamiltonian may take exponential time in $q$.

Finally, we show that the $\eps^{-2}$ sample dependence for product-state energy is optimal, even for diagonal Hamiltonians on qubits (\cref{cor:quantum-particles}). The same reduction gives an $\Omega_r(\eps^{-2})$ sampling lower bound for ground-state energy; we do not claim that the $\eps^{-3}$ upper bound is optimal.

\subsection{Technical Overview}\label{sec:overview}

We now explain the main ideas behind the bounds on the hindsight gain. We begin with why a direct uniform-convergence argument loses too much, and then describe the comparison argument that avoids this loss.

\paragraph{Uniform convergence is too lossy.}
A natural first attempt bounds the gain by
\[
\E\sup_{x\in K}\bigl[f_{A,\boldsymbol{\zeta}}(x)-f_{A,\mathbf{p}}(x)\bigr].
\]
This can lose a polynomial factor even for MaxCut. Take $G=K_n$ and $p=1/2$. With high probability, a cut one of whose sides consists of $n/6$ retained vertices keeps $(1/48-o(1))n^2$ more crossing edges than its expectation. Yet the hindsight gain is only
$
\E\left\lfloor\frac{\abs{S_p}^2}{4}\right\rfloor
-\frac14\left\lfloor\frac{n^2}{4}\right\rfloor
=\frac n{16}+O(1)
$.
The optimum is much more stable than the uniform deviation. Our argument compares the two optima directly.

\paragraph{Step 1: a fixed objective over a random convex body.}
Using exact convexification and the substitution $y_i=\zeta_i x_i$, we obtain
\begin{equation}\label{eq:body-intro}
\Val_A(\boldsymbol{\zeta})=\max_{y\in Z_{\boldsymbol{\zeta}}}f_A(y),
\qquad Z_{\boldsymbol{\zeta}}:=\prod_i\zeta_i\conv K_i.
\end{equation}
This transforms the problem from a random objective into a fixed multilinear objective over a random feasible body. 
Its support function, the largest linear score in a given direction, is
$h_{Z_{\boldsymbol{\zeta}}}(v)=\sum_i\zeta_i h_{K_i}(v_i)$.
This expression is linear in the independent weights. After this step, we turn its fluctuations into a comparison of maxima of $f_A$.

\paragraph{Step 2: a support direction for the difference of two maxima.}
Let $y$ maximize $f_A$ on $Z_{\boldsymbol{\zeta}}$. The deterministic comparison lemma in \cref{sec:comparison} constructs a point $z\in Z_{\mathbf{p}}$ and a direction $v$ such that
\[
f_A(y)-f_A(z)=\ip v{y-z},
\qquad h_{Z_{\mathbf{p}}}(v)=\ip vz.
\]
It chooses $z$ by maximizing an auxiliary polynomial on $Z_{\mathbf{p}}$; the first-order optimality condition makes $z$ a support point in the required direction. Consequently,
\begin{equation}\label{eq:comparison-intro}
\Val_A(\boldsymbol{\zeta})-\Val_A(\mathbf{p})
\le h_{Z_{\boldsymbol{\zeta}}}(v)-h_{Z_{\mathbf{p}}}(v).
\end{equation}
This comparison uses neither convexity nor a curvature bound for the objective.

\paragraph{Step 3: projection and vector contraction.}
We now bound the support-function fluctuation in~\eqref{eq:comparison-intro}. We write the comparison direction as $v=LW$, separating the fixed interaction coefficients in $L$ from the products of coordinates of $y$ and $z$ in $W$, with $\fro L\le\fro A$ (\cref{sec:representation}). To control the fluctuation uniformly over normalized arrays, we define
\[
\Delta:=\max_{\norm w\le1}\bigl[h_{Z_{\boldsymbol{\zeta}}}(Lw)-h_{Z_{\mathbf{p}}}(Lw)\bigr].
\]
Symmetrization and Maurer's vector-contraction inequality~\cite{Mau16} control its mean, while the Efron--Stein inequality controls its variance. Together, these give $\E\Delta^2\le5\sigma_{\max}^2\fro A^2$.

It remains to control the size of $W$. Since each interaction involves distinct variables, we can project away entries with repeated variable indices without changing $LW$. Independence then lets us factor the relevant expectations, so only second moments of the weights appear. We obtain a bound $R(\boldsymbol{\zeta})$ on the projected norm such that
\[
\Val_A(\boldsymbol{\zeta})-\Val_A(\mathbf{p})\le R(\boldsymbol{\zeta})\Delta,
\qquad
\bigl(\E R(\boldsymbol{\zeta})^2\bigr)^{1/2}\le rM^{(r-1)/2}.
\]
We now combine the size and fluctuation bounds by Cauchy--Schwarz:
\[
\Gap_A(\boldsymbol{\zeta})
\le\bigl(\E R(\boldsymbol{\zeta})^2\bigr)^{1/2}\bigl(\E\Delta^2\bigr)^{1/2}
\le r\sqrt5\,M^{(r-1)/2}\sigma_{\max}\fro A.
\]
This proves \cref{thm:noise}; we give the detailed estimates in \cref{sec:fluct,sec:proof-noise}.

\paragraph{Step 4: iteration with exact factorial moments.}
We next sharpen the bound for uniform retention by repeatedly halving the sampled instance. To track the change in the rescaled optimum, we set $F_A(t)=t^{-r}\E\Val_A(S_t)$. Write $A_{S_t}$ for the interactions retained on $S_t$. Halving the random instance on $S_t$ lets us express the increment $F_A(t/2)-F_A(t)$ through the bound from Steps~1--3. Here we keep a little more information from the projected norm estimate: on an instance with $m$ active variables, it retains the factor $\sqrt{(m)_{r-1}}$, where $(m)_{r-1}$ counts ordered tuples of distinct variables. Consequently,
\[
F_A(t/2)-F_A(t)
\le C_rt^{-r}\E\bigl[\sqrt{(\abs{S_t})_{r-1}}\,\fro{A_{S_t}}\bigr]
\le C_rt^{-1/2}n^{(r-1)/2}\fro A.
\]
The last inequality follows from Cauchy--Schwarz and the exact identities
$\E(\abs{S_t})_{r-1}=t^{r-1}(n)_{r-1}$ and
$\E\fro{A_{S_t}}^2=t^r\fro A^2$.
We then sum over dyadic scales down to $p$. The resulting geometric series is dominated by its last term, so multiplying by $p^r$ gives the desired $p^{r-1/2}$ factor. For $p\ge1/2$, a single step instead retains the factor $\sqrt{1-p}$. Together, these arguments prove \cref{thm:uniform} for all sampling rates.

\paragraph{Applications.}
We finish with the main ideas behind the sampling guarantees and the corresponding lower bounds. For Max-$r$CSPs, we use one-hot encoding and center each table to express the bias in terms of $\cE(g)$. To control the variance as well, we apply the Efron--Stein inequality, charging a variable only for interactions whose other variables were sampled. We then expand the squared charge and group pairs of constraints by their overlap size. After normalization, this gives fluctuations of order $O_{r}(n^r/\sqrt q)$, matching the bias for expected sample size $q$. The graph applications follow by specializing to $r=2$.

To establish the graph lower bound, we compare $G(n,1/2)$ with a planted model in which edges across a hidden random bipartition are slightly more likely. Every pair is an edge with probability $1/2$ under both models, so the distinguishing information lies in dependencies among the query answers. We express the $\chi^2$-divergence for a fixed query graph as a weighted count of its nonempty even subgraphs, with weight $\theta^{2j}$ for $j$ edges and $\theta\asymp\eps$. A spectral cycle count bounds this divergence, keeping it small for fewer than $c\eps^{-4}$ queries. We then use a random relabeling argument~\cite{GT03} to obtain an $\Omega(\eps^{-2})$ lower bound on examined vertices, even with adaptive queries. Finally, by padding graph constraints with variables whose labels do not affect their payoffs, the sampling lower bound can be extended to every fixed arity.

For the quantum application, we use the fact that product-state energies are multilinear in the single-particle density matrices, whose real coordinate vectors form compact subsets of the unit ball. Our argument therefore applies directly to these continuous domains. Coupling Bernoulli and fixed-size samples gives error $O_{r,d}(n^r/\sqrt q)$ for product-state energies. To pass to full ground-state energies, we then apply the known product-state approximation theorem. Its additional error of $O_{r,d}(n^r/q^{1/3})$ yields the sample bound of $O_{r,d}(\eps^{-3})$ particles.

\subsection{Related Work}\label{sec:related}

\paragraph{Value of information and optimization bias.}
Our hindsight gain is the expected value of perfect information in stochastic programming~\cite{AW70,BL11}. Huang, Vertinsky, and Ziemba~\cite{HVZ77} give sharp bounds for simple-recourse problems. The same bias appears in sample average approximation~\cite{MMW99}, and its reduction through smoothing is studied in~\cite{DL22}. Natarajan, Shi, and Toh~\cite{NST18} bound expected binary quadratic optima given marginal coefficient distributions. Our model instead specifies the joint law through independent variable weights.

\paragraph{Selection bias and sequential adaptivity.}
The optimizer's curse~\cite{SW06} and information-theoretic bounds for adaptive data analysis~\cite{RZ16} concern optimism after selecting noisy estimates. For a sampled maximizer, this optimism equals our optimal-value bias plus its expected regret under the mean objective. In stochastic knapsack, Dean, Goemans, and Vondr{\'a}k~\cite{DGV08} compare sequential adaptive and nonadaptive policies multiplicatively. Here the entire restriction is revealed before optimization, and the comparison is additive. Information relaxations of stochastic dynamic programs~\cite{BSS10} also use perfect-information benchmarks, with penalties for decisions that use future information.

\paragraph{Dense optimization and testing.}
The closest sampling bounds are those of Alon, Fernandez de la Vega, Kannan, and Karpinski~\cite{AFKK03}, Rudelson and Vershynin~\cite{RV07}, and Mathieu and Schudy~\cite{MS08}, discussed in \cref{sec:apps}. In our terminology, \cite[Theorem~1.4]{MS08} gives rescaled hindsight gain $O(n^2q^{-1/4})$ for a uniform sample of $q$ vertices; \cref{thm:uniform} gives $O(n^2q^{-1/2})$ for a Bernoulli sample of expected size $q$. Graph and CSP limits relate sampling to testability and convergence of cut and energy parameters~\cite{BCLSV08,BCLSV12,KM14}. Subsampling of LP and SDP relaxations~\cite{BHHS11} provides precedents for continuous domains; our theorem handles compact local sets directly through the Frobenius norm of the interaction coefficients. Core-sets extend MaxCut sampling beyond dense graphs~\cite{BDV18}. Ordinary and tolerant bipartiteness testing~\cite{AK02,BT04,GMRS22} ask for weaker distinctions than full additive estimation.

For growing arity or alphabet size, other parameter dependences become important. Barak, Hardt, Holenstein, and Steurer~\cite{BHHS11} obtain polylogarithmic alphabet dependence for fixed arity. Bernasconi, Castiglioni, Celli, Farina, and Malavolta~\cite{BCCFM26} give a sample bound polynomial in the arity and polylogarithmic in the alphabet size. Our bounds optimize the accuracy dependence for fixed arity and alphabet; the explicit $k^r$ factor is not intended to be optimal when these parameters grow.

\paragraph{Statistical physics and quantum Hamiltonians.}
Jain, Koehler, and Mossel~\cite{JKM18} prove polynomial vertex sample complexity for free energies of Ising models and higher-order Markov random fields. Jain, Koehler, and Risteski~\cite{JKR19} give sharp bounds on mean-field approximation error. Brand\~ao and Harrow~\cite{BH13} establish product-state approximations to quantum ground-state energies; Bergamaschi~\cite{Bergamaschi23} extends these guarantees and proves vertex-sampling results. \Cref{sec:quantum} combines our sharper product-state sampling bound with his approximation theorem to estimate the full ground-state energy.

\paragraph{Empirical and random processes.}
Pairwise empirical optimization also appears in the $U$-process analysis of ranking~\cite{CLV08}, where the decision is a prediction rule rather than a separate choice at each vertex. General chaos-process bounds, such as the quadratic bounds of~\cite{KMR14}, control a supremum that can be polynomially larger than our gain. Our comparison principle reduces the optimal-value difference to a process controlled by vector contraction~\cite{LT91,Mau16}.

\paragraph{Organization.}
\Cref{sec:prelim} introduces the notation and structural properties. \Cref{sec:onestep,sec:uniform} prove the general and uniform bounds on the hindsight gain. 
The following sections give the applications.
\Cref{sec:csp} discusses Max-$r$CSPs, MaxCut, and bipartiteness; \cref{sec:zhao,sec:quantum} give the percolation and quantum applications. The lower bounds appear in \cref{sec:lower}.

\section{Preliminaries}\label{sec:prelim}

We recall the model of \cref{sec:model}. Let $n,D\ge1$ be integers and fix an integer $r\ge2$. Consider a collection of variables $\{x_i\}_{i=1}^n$. Each variable $i$ chooses a vector $x_i\in K_i$, where $K_i\subseteq\R^D$ is a nonempty compact subset of the Euclidean unit ball. Note that $K_i$ need not be convex.

Write $[n]=\{1,\ldots,n\}$. For each $e=\{i_1<\cdots<i_r\}\in\binom{[n]}r$, let $A_e\colon(\R^D)^r\to\R$ be a multilinear function, meaning linear in each argument separately, and write $x_e=(x_{i_1},\ldots,x_{i_r})$. For $x=(x_1,\ldots,x_n)$, the objective is
\[
f_A(x):=\sum_{e\in\binom{[n]}r}A_e(x_e).
\]
Thus each interaction involves exactly $r$ distinct variables. Missing interactions are considered zero.

All vector norms and inner products are Euclidean. Constants denoted by $C_r,c_r,\ldots$ depend only on $r$; a subscript $d$ or $k$ permits dependence on the local dimension or alphabet size as well.

\paragraph{Coefficients.}
For $e=\{i_1<\cdots<i_r\}$, write the interaction in coordinates as
\[
A_e(x_e)=\sum_{a_1,\ldots,a_r=1}^D
(A_e)_{a_1,\ldots,a_r}\prod_{j=1}^r x_{i_j,a_j},
\qquad
\fro{A_e}^2=\sum_{a_1,\ldots,a_r=1}^D(A_e)_{a_1,\ldots,a_r}^2.
\]
For the collection $A=(A_e)_e$, define
\begin{equation}\label{eq:coefficient-norm}
\fro A^2:=\sum_{e\in\binom{[n]}r}\fro{A_e}^2,
\end{equation}
so that each interaction is counted once. These norms are unchanged by orthonormal changes of coordinates for the variables.

\paragraph{Optimal values.}
Write $K:=\prod_{i=1}^nK_i$. For nonnegative weights $a=(a_i)_{i=1}^n$, define the weighted objective and its optimum by
\[
f_{A,a}(x):=\sum_{e\in\binom{[n]}r}\Bigl(\prod_{i\in e}a_i\Bigr)A_e(x_e),
\qquad
\Val_A(a):=\max_{x\in K}f_{A,a}(x).
\]
The full and restricted optima are the special cases
\[
\Val_A:=\Val_A(\one)=\max_{x\in K}f_A(x),
\qquad
\Val_A(S):=\Val_A(\one_S)\quad(S\subseteq[n]),
\]
where $\one\in\R^n$ is the all-ones vector and $\one_S\in\R^n$ is the indicator vector of $S$. Thus $\Val_A(S)$ keeps only interactions entirely contained in $S$ and is zero when $\abs S<r$.

\paragraph{Hindsight gain.}
Let $\zeta_1,\ldots,\zeta_n$ be independent nonnegative weights with finite second moments and means $p_i$, and write $\boldsymbol{\zeta}=(\zeta_i)_i$ and $\mathbf{p}=(p_i)_i$. Define the \emph{hindsight gain} by
\begin{equation}\label{eq:gap-prelim}
\Gap_A(\boldsymbol{\zeta}):=\E\Val_A(\boldsymbol{\zeta})-\Val_A(\mathbf{p}).
\end{equation}
This measures the expected benefit of choosing an assignment after observing the weights, compared with choosing an assignment that maximizes the mean objective.

Note that independence gives $\E\prod_{i\in e}\zeta_i=\prod_{i\in e}p_i$, so $\E f_{A,\boldsymbol{\zeta}}(x)=f_{A,\mathbf{p}}(x)$ for every fixed $x\in K$. In other words, the random objective is an unbiased estimate of the mean objective.
Evaluating the random objective at a maximizer of the mean objective yields $\E\Val_A(\boldsymbol{\zeta})\ge\Val_A(\mathbf{p})$. Hence we always have $\Gap_A(\boldsymbol{\zeta})\ge0$.

\paragraph{Integrability.}
The weighted optimum $\Val_A(a)$ is continuous in $a$, by joint continuity of $f_{A,a}(x)$ and compactness of $K$. Thus $\Val_A(\boldsymbol{\zeta})$ is measurable. Moreover, Cauchy--Schwarz applied to the interaction coefficients gives
$\abs{\Val_A(a)}\le\fro A
\left(\sum_{e\in\binom{[n]}r}\prod_{i\in e}a_i^2\right)^{1/2}$.
Squaring it and using independence of the weights yields
\[
\E\abs{\Val_A(\boldsymbol{\zeta})}^2
\le\fro A^2\sum_{e\in\binom{[n]}r}\prod_{i\in e}\E\zeta_i^2
\le\fro A^2\left(\sum_i\E\zeta_i^2\right)^r<\infty.
\]
Hence independent weights with finite second moments give a square-integrable weighted optimum, so the expectation defining the hindsight gain is finite.

\paragraph{Convexification.}
To compare weighted optima, we move the weights from the objective into a convex feasible set without changing the optimal value.
Scaling each vector $x_i$ by $a_i$ incorporates the weight of every interaction, so
$f_{A,a}(x)=f_A(a_1x_1,\ldots,a_nx_n)$. For $a\in\R^n_{\ge0}$ put
\begin{equation}\label{eq:body}
Z_a:=\prod_{i=1}^n a_i\conv K_i.
\end{equation}
Here $\conv K_i$ denotes the convex hull of $K_i$. The set $Z_a$ is a nonempty compact convex subset of $\R^{nD}$. The following lemma lets us use it while keeping arbitrary compact local sets in the model.

\begin{lemma}[Exact convexification]\label{lem:affinity}
For every $a\in\R^n_{\ge0}$,
\begin{equation}\label{eq:convexification}
\Val_A(a)=\max_{y\in Z_a}f_A(y).
\end{equation}
\end{lemma}

\begin{proof}
The objective is affine in each vector variable separately. Given $x_i\in\conv K_i$, choose independent, finitely supported $K_i$-valued random vectors $X_i$ with $\E X_i=x_i$. Since every interaction uses distinct variables, $\E f_{A,a}(X)=f_{A,a}(x)$, so some realization in $K$ attains at least this value. Thus maximizing $f_{A,a}$ over $\prod_i\conv K_i$ gives the same value as maximizing over $K$, since $K_i\subseteq\conv K_i$. Every $y\in Z_a$ can be written as $y_i=a_ix_i$ with $x_i\in\conv K_i$, and $f_A(y)=f_{A,a}(x)$. Conversely, every such $x$ gives a point $y\in Z_a$, proving~\eqref{eq:convexification}.
\end{proof}

\paragraph{Support functions.}
To compare the feasible sets obtained above, we will use their support functions.
For a compact set $B$, its support function
$h_B(v):=\max_{y\in B}\ip yv$ is the largest linear score in direction $v$. It is positively homogeneous and satisfies $h_{\conv B}=h_B$. Since $\norm{x_i}\le1$ on $K_i$, each $h_{K_i}$ is $1$-Lipschitz and $\abs{h_{K_i}(v)}\le\norm v$. Maximizing independently over the variables gives
\begin{equation}\label{eq:support-product}
h_{Z_a}(v)=\sum_{i=1}^n a_i h_{K_i}(v_i),
\qquad v=(v_1,\ldots,v_n)\in\R^{nD}.
\end{equation}

\section{General Hindsight Gain}\label{sec:onestep}

In this section, we prove \cref{thm:noise}, which bounds the hindsight gain under independent random weights. We first compare maxima of polynomials over two convex bodies through a single support direction, then use a vector-contraction inequality to control the fluctuations of the random body's support function. The comparison direction is a linear image of an array of coordinate products. By removing repeated variable indices before estimating its norm, we will need only second moments of the weights.

We use the following notation throughout this section:
\[
p_i=\E\zeta_i,\qquad M=\sum_i\E\zeta_i^2,\qquad
\bar M=\sum_i p_i^2\le M,\qquad \sigma_{\max}=\max_i\sqrt{\Var(\zeta_i)}.
\]

\subsection{A Coordinate Representation for the Proof}\label{sec:representation}

To prepare for the comparison argument, we collect the vectors into $x=(x_1,\ldots,x_n)\in E:=\R^{nD}$ and represent the homogeneous polynomial $f_A$ by a symmetric $r$-linear form. Let $\sigma$ be a uniformly random permutation of $[r]$. For $u^1,\ldots,u^r\in E$, define
\begin{equation}\label{eq:symmetrization}
T(u^1,\ldots,u^r):=\E_\sigma
\sum_{e=\{i_1<\cdots<i_r\}}
A_e(u^{\sigma(1)}_{i_1},\ldots,u^{\sigma(r)}_{i_r}),
\end{equation}
where the sum is over $r$-element subsets of $[n]$. This average is symmetric in its arguments, and $T(x,\ldots,x)=f_A(x)$. For each fixed permutation, the form inside the average has the same coefficients as $A$, placed in distinct positions. Averaging cannot increase the Euclidean norm of the coefficient array, so
\begin{equation}\label{eq:symmetrized-norm}
\fro T\le\fro A.
\end{equation}
This representation is used only in the proof; the interactions $A_e$ themselves need no symmetry. We abbreviate repeated arguments by $T(y^k,z^{r-k})$, and write $f_T(x)=T(x,\ldots,x)$ for a general symmetric form.

We use tensor notation to keep track of coordinate products. For vectors $v^1,\ldots,v^s\in E$, the tensor product $v^1\otimes\cdots\otimes v^s$ is the array whose entries are products of their coordinates. We give the full array space $E^{\otimes s}$ its Euclidean inner product, so that
\[
\ip{u^1\otimes\cdots\otimes u^s}{v^1\otimes\cdots\otimes v^s}
=\prod_{j=1}^s\ip{u^j}{v^j}.
\]
Write $y^{\otimes s}$ for a repeated product. To separate the interaction coefficients from these products, we reshape the coefficients of $T$ into a matrix $L\colon E^{\otimes(r-1)}\to E$, with the first coordinate as its row index:
\begin{equation}\label{eq:flattening}
\ip{u}{L(v^1\otimes\cdots\otimes v^{r-1})}=T(u,v^1,\ldots,v^{r-1}).
\end{equation}
Let $L_i$ consist of the $D$ rows belonging to variable $i$, so $L_iw=(Lw)_i$. For any matrix $B$, write $\opn B$ and $\fro B=(\tr B^*B)^{1/2}$ for its operator and Frobenius norms; in particular, $\opn B\le\fro B$. The norm bound above and the same averaging argument for the rows give
\begin{equation}\label{eq:row-norms}
\fro L^2=\sum_i\fro{L_i}^2\le\fro A^2,
\qquad
\fro{L_i}^2\le\frac1r\sum_{e\ni i}\fro{A_e}^2.
\end{equation}
To see the second inequality, consider an interaction containing $i$: a uniformly random permutation places that variable in the first argument with probability $1/r$. We then apply convexity of the squared Frobenius norm to the corresponding rows.

One more feature of this representation will let us work with only second moments. Every coordinate of $E$ has a variable index $i$ and an internal coordinate index $a\in[D]$. Let $P_{\ne}$ be the coordinate projection that keeps entries of $E^{\otimes(r-1)}$ only when their variable indices are pairwise distinct. Because the interactions use distinct variables, the other columns of $L$ are zero, and hence
\begin{equation}\label{eq:projection}
L=LP_{\ne}.
\end{equation}
This identity lets us remove repeated variable indices before estimating the norm of the comparison direction.

\subsection{A Comparison Principle}\label{sec:comparison}

Our next step is to bound the difference of two maxima by a difference of support functions in one direction. The following deterministic lemma does this without a sign or convexity assumption on the objective.

\begin{lemma}[Comparison principle]\label{lem:comparison}
Let $T$ be a symmetric $r$-linear form on a Euclidean space $E$, and let $L$ be its linear representation~\eqref{eq:flattening}. Let $Z\subseteq E$ be nonempty and compact, and let $Z'\subseteq E$ be nonempty, compact, and convex. If $y$ maximizes $f_T$ over $Z$, then there is $z\in Z'$ such that, with
\begin{equation}\label{eq:comparison-direction}
W(y,z):=\sum_{k=0}^{r-1}y^{\otimes k}\otimes z^{\otimes(r-1-k)},
\end{equation}
we have
\begin{equation}\label{eq:comparison}
\max_Z f_T-\max_{Z'}f_T
\le h_Z(LW(y,z))-h_{Z'}(LW(y,z)).
\end{equation}
The zeroth tensor power in~\eqref{eq:comparison-direction} is the scalar $1$.
\end{lemma}

\begin{proof}
We choose the reference point $z\in Z'$ by maximizing the auxiliary polynomial
\[
\Phi_y(u):=\sum_{j=1}^r\frac1j T(y^{r-j},u^j).
\]
Symmetry of $T$ and differentiation give $\nabla\Phi_y(z)=LW(y,z)$. Since the segment from $z$ to any $u\in Z'$ lies in $Z'$, the first-order necessary condition at this maximum yields
\[
\ip{LW(y,z)}{u-z}\le0\qquad(u\in Z').
\]
Hence $h_{Z'}(LW)=\ip{LW}{z}$. To connect this support direction to the difference of objective values, we use the telescoping identity supplied by multilinearity:
\[
f_T(y)-f_T(z)
=\sum_{k=0}^{r-1}T(y^k,z^{r-1-k},y-z)
=\ip{LW}{y-z}.
\]
Since $y$ maximizes the objective on $Z$, we can now conclude that
\[
\max_Zf_T-\max_{Z'}f_T
\le f_T(y)-f_T(z)
\le h_Z(LW)-h_{Z'}(LW),
\]
as claimed.
\end{proof}

\begin{figure}[htbp]
\definecolor{bodyblue}{HTML}{295D94}
\definecolor{refgreen}{HTML}{52724A}
\definecolor{valueorange}{HTML}{AD531E}
\centering
\begin{tikzpicture}[x=1.1cm,y=0.95cm,>=Stealth,font=\small,
  every node/.style={inner sep=2pt},
  support/.style={densely dashed,line width=0.5pt},
  point/.style={circle,fill=black,inner sep=1.4pt}]
  \fill[bodyblue,fill opacity=0.07] (6.4,4.15) ellipse[x radius=3,y radius=1.3];
  \fill[refgreen,fill opacity=0.10] (3,4) ellipse[x radius=1.55,y radius=1];
  \draw[bodyblue,line width=0.8pt] (6.4,4.15) ellipse[x radius=3,y radius=1.3];
  \draw[refgreen,line width=0.8pt] (3,4) ellipse[x radius=1.55,y radius=1];
  \node[text=refgreen] at (2.35,4) {$Z'$};
  \node[text=bodyblue] at (8.3,3.45) {$Z$};

  \coordinate (z) at (4.55,4);
  \coordinate (y) at (7.6,{4.15+1.3*sqrt(1-(1.2/3)^2)});
  \coordinate (xv) at (9.4,4.15);

  \draw[support,refgreen] (4.55,5.6) -- (4.55,-0.1);
  \draw[support,bodyblue] (9.4,5.6) -- (9.4,-0.1);
  \draw[densely dotted,valueorange,line width=0.6pt] (y) -- (7.6,0.9);

  \draw[->,valueorange,line width=0.9pt] (z) -- (y)
    node[pos=0.55,above=2pt,sloped] {$y-z$};
  \draw[->,line width=0.9pt] (z) -- (6.3,4) node[right] {$v$};
  \node[point,label={[label distance=1pt]left:$z$}] at (z) {};
  \node[point,label={above:$y$}] at (y) {};
  \node[point,label={right:$x_v$}] at (xv) {};

  \draw[->,black!65,line width=0.65pt] (1.2,1.9) -- (10.3,1.9)
    node[right,text=black] {$\ip{v}{x}$};
  \foreach \p in {4.55,7.6,9.4}
    \draw[black!65,line width=0.6pt] (\p,1.99) -- (\p,1.81);
  \node[anchor=north,fill=white,text=refgreen] at (4.55,1.73)
    {$h_{Z'}(v)=\ip{v}{z}$};
  \node[anchor=north,fill=white,text=valueorange] at (7.6,1.73)
    {$\ip{v}{y}$};
  \node[anchor=north,fill=white,text=bodyblue] at (9.4,1.73)
    {$h_Z(v)$};

  \draw[<->,valueorange,line width=1pt] (4.55,0.9) -- (7.6,0.9)
    node[midway,below=4pt,fill=white] {$f_T(y)-f_T(z)$};
  \draw[<->,bodyblue,line width=1pt] (4.55,-0.1) -- (9.4,-0.1)
    node[midway,below=4pt,fill=white] {$h_Z(v)-h_{Z'}(v)$};
\end{tikzpicture}
\caption{Comparison through linear scores in direction $v=\nabla\Phi_y(z)=LW(y,z)$, drawn for $v\ne0$ and positive differences. Here $y$ maximizes $f_T$ on $Z$, the auxiliary maximizer $z$ of $\Phi_y$ supports $Z'$, and $x_v$ supports $Z$. The orange interval $f_T(y)-f_T(z)=\ip{v}{y-z}$ lies inside the blue interval $h_Z(v)-h_{Z'}(v)$. Since $\max_{Z'}f_T\ge f_T(z)$, the difference of optima is at most the orange interval.}
\label{fig:comparison-support}
\end{figure}
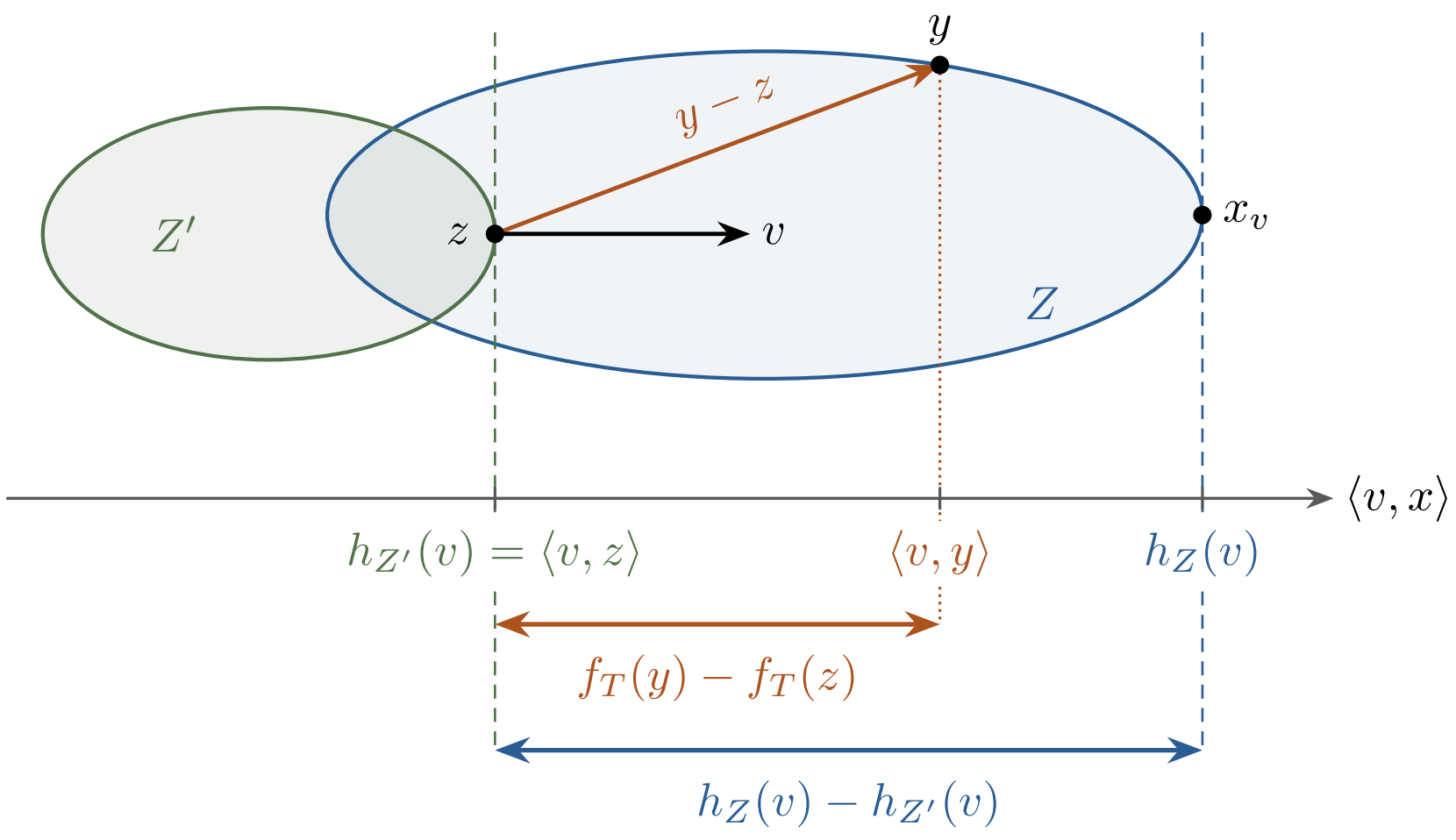

Only the reference body $Z'$ needs to be convex. We use the auxiliary polynomial to make $z$ a support point in exactly the direction that represents the difference of objective values; \cref{fig:comparison-support} illustrates this choice.

\subsection{Fluctuations of the Support Function}\label{sec:fluct}

The comparison lemma reduces our task to controlling a difference of support functions. We now estimate its fluctuations under the random weights.

By~\eqref{eq:support-product},
\begin{equation}\label{eq:support-diff}
h_{Z_{\boldsymbol{\zeta}}}(v)-h_{Z_{\mathbf{p}}}(v)
=\sum_i(\zeta_i-p_i)h_{K_i}(v_i).
\end{equation}
The comparison direction lies in the image of $L$, so we first control this difference on the linear image of a unit ball. Define
\begin{equation}\label{eq:delta}
\Delta(\boldsymbol{\zeta}):=\max_{\norm w\le1}\sum_i(\zeta_i-p_i)h_{K_i}(L_iw),
\qquad
V^2:=\sum_i\Var(\zeta_i)\fro{L_i}^2,
\end{equation}
where $w\in E^{\otimes(r-1)}$. The maximum is continuous in $\boldsymbol{\zeta}$ and is nonnegative, since $w=0$ is feasible. The bound
\[
\Delta(\boldsymbol{\zeta})\le\sum_i\abs{\zeta_i-p_i}\fro{L_i}
\]
also shows that it is square-integrable under the finite-second-moment assumption. To bound its mean, we use the following vector-contraction inequality.

\begin{theorem}[{Maurer~\cite[Theorem~3, preprint]{Mau16}}]\label{thm:maurer}
Let $\cS$ be countable, and let $\psi_i\colon\cS\to\R$ and $\phi_i\colon\cS\to\ell_2$ satisfy
$\psi_i(s)-\psi_i(s')\le\norm{\phi_i(s)-\phi_i(s')}$ for all $s,s'\in\cS$. For independent Rademacher sequences $(\epsilon_i)$ and $(\epsilon_{ik})$,
\[
\E\sup_{s\in\cS}\sum_i\epsilon_i\psi_i(s)
\le\sqrt2\,\E\sup_{s\in\cS}\sum_{i,k}\epsilon_{ik}\phi_i(s)_k.
\]
\end{theorem}

\begin{lemma}[Fluctuation bound]\label{lem:fluct}
The quantities in~\eqref{eq:delta} satisfy $\E\Delta\le2V$, $\Var\Delta\le V^2$, and hence $\E\Delta^2\le5V^2$.
\end{lemma}

\begin{proof}
We first bound the mean. Write $X_i(w)=h_{K_i}(L_iw)$, and let $\boldsymbol{\zeta}'$ be an independent copy of $\boldsymbol{\zeta}$. Symmetrization gives
\[
\E\Delta\le\E\max_{\norm w\le1}\sum_i(\zeta_i-\zeta_i')X_i(w)
=\E\max_{\norm w\le1}\sum_i\epsilon_i(\zeta_i-\zeta_i')X_i(w),
\]
where the last equality uses independence and symmetry of the differences. Conditional on $(\boldsymbol{\zeta},\boldsymbol{\zeta}')$, put $a_i=\zeta_i-\zeta_i'$. On a countable dense subset of the unit ball, apply \cref{thm:maurer} with $\psi_i(w)=a_iX_i(w)$ and $\phi_i(w)=a_iL_iw$; its hypothesis follows from the $1$-Lipschitz property of $h_{K_i}$. Continuity allows us to take the supremum over the entire ball. If $\epsilon^{(i)}$ is a vector of independent signs in $\R^D$, we obtain
\begin{align*}
\E_\epsilon\max_{\norm w\le1}\sum_i\epsilon_i a_iX_i(w)
&\le\sqrt2\,\E\left\|\sum_i a_iL_i^*\epsilon^{(i)}\right\|\\
&\le\sqrt2\left(\sum_i a_i^2\fro{L_i}^2\right)^{1/2}.
\end{align*}
The last line follows from independence of the signs and the Cauchy--Schwarz inequality. Averaging over the two weight vectors and using $\E a_i^2=2\Var(\zeta_i)$ proves $\E\Delta\le2V$.

For the variance, we examine the effect of changing one weight. Since $\abs{X_i(w)}\le\fro{L_i}$ on the unit ball, replacing $\zeta_i$ by an independent copy changes $\Delta$ by at most $\abs{\zeta_i-\zeta_i'}\fro{L_i}$. The Efron--Stein inequality~\cite{BLM13} therefore gives
\[
\Var\Delta\le\frac12\sum_i\E(\zeta_i-\zeta_i')^2\fro{L_i}^2=V^2.
\]
Combining the two estimates proves the last assertion.
\end{proof}

\subsection{The General Bound}\label{sec:proof-noise}

We now combine the comparison principle with the fluctuation bound. The remaining step is to control the size of the projected comparison direction using only second moments.

\begin{theorem}[Refined bound]\label{thm:refined}
In the model of \cref{sec:model}, put $b_i^2:=\sum_{e\ni i}\fro{A_e}^2$, the sum of squared coefficients of the interactions involving variable $i$. Then
\begin{equation}\label{eq:refined}
0\le\Gap_A(\boldsymbol{\zeta})
\le\sqrt{\frac5r}\left(\sum_i\Var(\zeta_i)b_i^2\right)^{1/2}
\sum_{k=0}^{r-1}M^{k/2}\bar M^{(r-1-k)/2}.
\end{equation}
\end{theorem}

\begin{proof}
Nonnegativity follows from unbiasedness, as shown in \cref{sec:prelim}. For the upper bound, we fix a realization of the weights and apply \cref{lem:comparison} to the form $T$ in~\eqref{eq:symmetrization}, with $Z=Z_{\boldsymbol{\zeta}}$ and $Z'=Z_{\mathbf{p}}$. Since $f_T=f_A$, \cref{lem:affinity} identifies these maxima with the weighted optima. Let $y\in Z_{\boldsymbol{\zeta}}$, $z\in Z_{\mathbf{p}}$, and $W=W(y,z)$ be the points and tensor supplied by the comparison lemma. Using~\eqref{eq:projection}, \eqref{eq:support-diff}, and positive homogeneity, we separate the size of the direction from the fluctuation:
\begin{equation}\label{eq:pointwise}
\Val_A(\boldsymbol{\zeta})-\Val_A(\mathbf{p})\le\norm{P_{\ne}W}\,\Delta(\boldsymbol{\zeta}).
\end{equation}
To use our fluctuation bound, we need a second-moment estimate for the projected norm. We obtain it by bounding each term before taking expectations. Since $\norm{y_i}\le\zeta_i$ and $\norm{z_i}\le p_i$,
\begin{align}
\left\|P_{\ne}\bigl(y^{\otimes k}\otimes z^{\otimes(r-1-k)}\bigr)\right\|^2
&=\sum_{\substack{i_1,\ldots,i_{r-1}\\\text{pairwise distinct}}}
\prod_{j=1}^k\norm{y_{i_j}}^2
\prod_{j=k+1}^{r-1}\norm{z_{i_j}}^2\notag\\
&\le\sum_{\substack{i_1,\ldots,i_{r-1}\\\text{pairwise distinct}}}
\prod_{j=1}^k\zeta_{i_j}^2\prod_{j=k+1}^{r-1}p_{i_j}^2
=:R_k(\boldsymbol{\zeta})^2.\label{eq:projected-radius}
\end{align}
Thus $\norm{P_{\ne}W}\le\sum_{k=0}^{r-1}R_k(\boldsymbol{\zeta})$. Each $R_k$ is an explicit measurable function of the weights, so no measurable selection of $y$ or $z$ is needed. Independence of the distinct weights in every summand gives
\[
\E R_k^2\le M^k\bar M^{r-1-k}.
\]
Minkowski's inequality combines these estimates into a bound for the sum:
\[
\left(\E\left[\left(\sum_{k=0}^{r-1}R_k\right)^2\right]\right)^{1/2}
\le\sum_{k=0}^{r-1}M^{k/2}\bar M^{(r-1-k)/2}.
\]
We now substitute $\norm{P_{\ne}W}\le\sum_kR_k$ into~\eqref{eq:pointwise} and take expectations. The Cauchy--Schwarz inequality and \cref{lem:fluct} then prove~\eqref{eq:refined}, using $\fro{L_i}^2\le b_i^2/r$ from~\eqref{eq:row-norms}.
\end{proof}

The projection in~\eqref{eq:projected-radius} is essential: it leaves products of second moments of distinct weights, even though the optimizing points may depend on all the weights. No higher moment of an individual weight appears.

\ThmNoise*

\begin{proof}
To recover the bound in the introduction, we simplify~\eqref{eq:refined} using $\bar M\le M$ and
$\sum_i\Var(\zeta_i)b_i^2\le \sigma_{\max}^2\sum_i b_i^2=r \sigma_{\max}^2\fro A^2$.
\end{proof}

Finally, we specialize the weights to Bernoulli variables to obtain the nonuniform restriction bound stated in the introduction.

\begin{corollary}[Nonuniform retention]\label{cor:bernoulli}
If $S$ retains variable $i$ independently with probability $p_i$, then
\begin{equation}\label{eq:bern-nonuniform}
0\le\E\Val_A(S)-\Val_A(\mathbf{p})
\le r\sqrt5\,\fro A\left(\sum_i p_i\right)^{(r-1)/2}
\sqrt{\max_i p_i(1-p_i)}.
\end{equation}
\end{corollary}

\begin{proof}
Apply \cref{thm:noise} with $\zeta_i=\one[i\in S]$, so that $M=\sum_i p_i$ and $\sigma_{\max}^2=\max_i p_i(1-p_i)$.
\end{proof}

\section{Uniform Hindsight Gain}\label{sec:uniform}

Our aim in this section is to prove \cref{thm:uniform}. A direct application of \cref{thm:noise} to Bernoulli weights gives an upper bound proportional to $p^{r/2}$; in this section we obtain the stronger factor $p^{r-1/2}$ by restricting the instance in stages, retaining each remaining variable independently with probability at least $1/2$ at each stage. The distinct-index projection in the proof of \cref{thm:refined} is essential here: it gives a falling factorial of the number of active variables, whose expectation has the exact scaling needed for the iteration.

\subsection{Iterated Thinning}\label{sec:thinning}

We begin by recalling the restricted optimum: for $S\subseteq[n]$,
\[
\Val_A(S)=\Val_A(\one_S)
=\max_{x\in K}\sum_{e\in\binom Sr}A_e(x_e),
\]
with value zero when $\abs S<r$. Let $A_S$ retain $A_e$ when $e\subseteq S$ and replace it by zero otherwise. Since $S_t$ retains each variable independently with probability $t$, we have
\begin{equation}\label{eq:restricted-frobenius}
\fro{A_S}^2=\sum_{e\in\binom Sr}\fro{A_e}^2,
\qquad
\E\fro{A_{S_t}}^2=t^r\fro A^2.
\end{equation}
The expectation identity holds because each interaction survives with probability $t^r$.

For $s\in[0,1]$, an \emph{$s$-thinning} of a set retains its elements independently with probability $s$. If we first sample $S_t$ and then, conditionally on $S_t$, thin at rate $s$, the result has the law of $S_{st}$. For nonnegative integers $m,k$, write $(m)_k=m(m-1)\cdots(m-k+1)$ for the falling factorial, with $(m)_0=1$ and $(m)_k=0$ when $m<k$. It counts ordered $k$-tuples of distinct elements from an $m$-element set.

We first bound the gain from thinning a fixed set. Keeping the falling factorial in the projected-radius estimate from \cref{sec:onestep} will let us average over a random starting set in the next step.

\begin{corollary}[Thinning a fixed set]\label{cor:thinning}
If $s\in[0,1]$ and $U_s$ is an $s$-thinning of a set $U\subseteq[n]$, then
\begin{equation}\label{eq:bernoulli-factorial}
0\le\E\Val_A(U_s)-s^r\Val_A(U)
\le r\sqrt5\,s^{r/2}\sqrt{1-s}\,\sqrt{(\abs U)_{r-1}}\,\fro{A_U}.
\end{equation}
\end{corollary}

\begin{proof}
We apply the argument of \cref{thm:refined} to the restricted interactions $A_U$ and the weights $\zeta_i=\one[i\in U_s]$. The random weighted optimum is $\Val_A(U_s)$, and the mean-weight optimum is $s^r\Val_A(U)$. Since the weights and their means vanish outside $U$, only the $(\abs U)_{r-1}$ tuples of distinct indices in $U$ contribute to the explicit sum in~\eqref{eq:projected-radius}, giving
\[
\E R_k^2=(\abs U)_{r-1}s^{2(r-1)-k}.
\]
To turn this radius estimate into a bound on the gain, we combine the Cauchy--Schwarz and Minkowski inequalities with \cref{lem:fluct}:
\[
\E\Val_A(U_s)-s^r\Val_A(U)
\le\sqrt5\,V\sum_{k=0}^{r-1}(\E R_k^2)^{1/2}.
\]
For $0<s\le1$, each square root in the sum is at most $\sqrt{(\abs U)_{r-1}}\,s^{(r-1)/2}$. Combining this with $V\le\sqrt{s(1-s)}\,\fro{A_U}$ from~\eqref{eq:row-norms} proves~\eqref{eq:bernoulli-factorial}. At $s=0$ both sides vanish.
\end{proof}

To track the gain across successive restrictions, we normalize the expected optimum by its natural scaling factor. For $t\in(0,1]$, define the \emph{compensated optimum}
\begin{equation}\label{eq:FA}
F_A(t):=t^{-r}\E\Val_A(S_t).
\end{equation}
Thus $F_A(1)=\Val_A$ and $F_A(t)-\Val_A=t^{-r}\Gap_A(\one_{S_t})$. We now bound the change in this quantity after one further thinning.

\begin{lemma}[One thinning step]\label{lem:increment}
For $s,t\in(0,1]$,
\[
0\le F_A(st)-F_A(t)
\le r\sqrt5\,\frac{\sqrt{1-s}}{s^{r/2}}\,
t^{-1/2}\sqrt{(n)_{r-1}}\,\fro{A}.
\]
\end{lemma}

\begin{proof}
We couple the two sampling rates by taking $S$ to be a $t$-thinning of $[n]$ and, conditionally on $S$, taking $S'$ to be an $s$-thinning of $S$. Then $S'$ has the same law as $S_{st}$. Applying \eqref{eq:bernoulli-factorial} with $U=S$, conditionally on $S$, gives
\[
0\le\E[\Val_A(S')\mid S]-s^r\Val_A(S)
\le r\sqrt5\,s^{r/2}\sqrt{1-s}\,
\sqrt{(\abs S)_{r-1}}\,\fro{A_S}.
\]
To average this bound over $S$, we need to control the product on the right. Its two factors need not be independent, so we use the Cauchy--Schwarz inequality:
\[
\E\bigl[\sqrt{(\abs S)_{r-1}}\,\fro{A_S}\bigr]
\le\sqrt{\E(\abs S)_{r-1}\cdot\E\fro{A_S}^2}.
\]
The first expectation counts ordered $(r-1)$-tuples of distinct retained variables, so it equals $t^{r-1}(n)_{r-1}$. Each interaction involves $r$ distinct variables, and hence
\[
\E\fro{A_S}^2=t^r\fro{A}^2.
\]
Consequently,
\[
\E\bigl[\sqrt{(\abs S)_{r-1}}\,\fro{A_S}\bigr]
\le t^{r-1/2}\sqrt{(n)_{r-1}}\,\fro{A}.
\]
We substitute this estimate into the expectation of the conditional bound and divide by $(st)^r$ to prove the lemma.
\end{proof}

In particular, $F_A$ is nonincreasing. The key point for our iteration is that the factor $t^{-1/2}$ does not depend on the degree: the projected comparison direction contributes $t^{(r-1)/2}$, the restricted coefficient norm contributes $t^{r/2}$, and the compensation contributes $t^{-r}$. The factorial-moment calculation is exact even when $tn<1$.

\ThmUniform*

\begin{proof}
The lower bound follows by restricting an optimizer of the full instance. For the upper bound, we sum the one-step estimate along dyadic scales. Let $\ell\ge0$ be the integer with $2^{-\ell}\le p<2^{1-\ell}$, and set $t_j=2^{-j}$ for $0\le j<\ell$ and $t_\ell=p$.

The ratios $s_j=t_{j+1}/t_j$ lie in $[1/2,1]$ and their product is $p$, so $s_j\ge p$ and $r\sqrt5\,s_j^{-r/2}\sqrt{1-s_j}\le r\sqrt5\,2^{r/2}\sqrt{1-p}$. Writing $B:=n^{(r-1)/2}\fro{A}\ge\sqrt{(n)_{r-1}}\,\fro A$, we telescope \cref{lem:increment} to obtain
\[
F_A(p)-F_A(1)
\le r\sqrt5\,2^{r/2}\sqrt{1-p}\,B\sum_{j=0}^{\ell-1}2^{j/2}
\le\frac{r\sqrt5\,2^{r/2}}{\sqrt2-1}\,2^{\ell/2}\sqrt{1-p}\,B
\le C_r\,p^{-1/2}\sqrt{1-p}\,B,
\]
where the last step uses $2^\ell<2/p$ and $\sqrt{10}/(\sqrt2-1)<8$, so $C_r=8r\,2^{r/2}$ suffices. Since $r\ge2$, this constant is at most $8^r$. Finally, multiplying by $p^r$ returns us to the unnormalized gain and proves the theorem.
\end{proof}

\subsection{Sharpness}\label{sec:sharp}

We now show that scalar instances attain the dependence on the retention rate and on the number of variables. 
In other words, the upper bound in \cref{thm:uniform} is tight up to a constant factor that depends only on the degree.
One group of variables adapts to a balanced sum that becomes unbalanced after restriction; the remaining groups supply the other $r-2$ factors of the objective.

\begin{proposition}[Sharpness]\label{prop:sharp}
For every $r\ge2$ there is $c_r>0$ such that, for every fixed $p\in(0,1)$ and all sufficiently large $n$, there is a collection $A$ of $r$-variable interactions with scalar variables ($D=1$), with each $K_i$ equal to $\{1\}$ or $\{-1,1\}$, satisfying
\[
\E\Val_A(S_p)-p^r\Val_A
\ge c_r p^{r-1/2}\sqrt{1-p}\,
n^{(r-1)/2}\fro{A}.
\]
\end{proposition}

\begin{proof}
Let $m=2\lfloor n/(2r)\rfloor$, and partition $rm$ variables into groups $G_1,\ldots,G_r$ of size $m$. Any remaining variables have $K_i=\{1\}$ and no interactions. To allow adaptation only in the first group, we take $K_i=\{-1,1\}$ on $G_1$ and $K_i=\{1\}$ on the other groups. Choose signs $a_i\in\{-1,1\}$ on $G_2$ with $\sum_{i\in G_2}a_i=0$, and define $A$ by the objective
\[
f_A(x)
=\Bigl(\sum_{i\in G_1}x_i\Bigr)
 \Bigl(\sum_{i\in G_2}a_ix_i\Bigr)
 \prod_{j=3}^{r}\Bigl(\sum_{i\in G_j}x_i\Bigr),
\]
where an empty product equals one. Each monomial uses one variable from each group and defines one interaction with coefficient $1$ or $-1$. There are $m^r$ such interactions, so
\[
\fro{A}^2=m^r.
\]

The full objective vanishes because its second factor is zero, so $\Val_A=0$. To see what changes under restriction, put $U_j=\abs{G_j\cap S_p}$ and $W=\sum_{i\in G_2}a_i\one[i\in S_p]$. Maximizing over the retained variables in $G_1$ and using independence across groups gives
\[
\Val_A(S_p)=U_1\abs W\prod_{j=3}^rU_j,
\qquad
\E\Val_A(S_p)=(mp)^{r-1}\E\abs W.
\]
It remains to lower-bound $\E\abs W$, which we do by comparing its second and fourth moments.
Let $v=mp(1-p)$, and take $n$ large enough that $v\ge1$. Since $W=\sum_{i\in G_2}a_i(\one[i\in S_p]-p)$,
\[
\E W^2=v,
\qquad
\E W^4\le v+3v^2\le4v^2.
\]
H\"older's inequality turns these moment estimates into the bound we need:
\[
\E\abs W\ge
\frac{(\E W^2)^{3/2}}{(\E W^4)^{1/2}}
\ge\frac12\sqrt v.
\]
Therefore the gain is at least
\[
\frac12m^{r-1/2}p^{r-1/2}\sqrt{1-p}.
\]
To express this gain at the scale in the proposition, note that for $m\ge2$, $n<r(m+2)\le2rm$, and hence
\[
n^{(r-1)/2}\fro{A}
\le(2r)^{(r-1)/2}\,m^{r-1/2}.
\]
The claim follows, for example, with $c_r=1/[2(2r)^{(r-1)/2}]$.
\end{proof}

\section{Application I: Max-\texorpdfstring{$r$}{r}CSPs, MaxCut, and Bipartiteness}\label{sec:csp}

A Max-$r$CSP assigns a label to each variable and collects a payoff from every constraint on $r$ variables. In this application, we use our uniform-retention theorem (\cref{thm:uniform}) to estimate its optimum from the constraints among a small random sample of variables, and then specialize to MaxCut and the distance to bipartiteness.

Given a set $S$ that retains each variable independently with probability $p$, we estimate the optimum $\Opt_g$ by the rescaled sampled optimum $p^{-r}\Opt_g(S)$. For payoffs in $[0,1]$, fixed $r$, and a fixed bound on the alphabet sizes, we will show that an expected sample size of $O(\eps^{-2})$ suffices for additive error $\eps n^r$ with probability at least $2/3$. This is the guarantee of \cref{thm:csp-intro}. For graphs, the same estimator, together with an estimate of the number of edges, gives part~\ref{it:bip-upper} of \cref{thm:bip-intro}.

Our proof bounds the bias and the fluctuations of the estimator separately. In \cref{sec:encoding}, we write a Max-$r$CSP as a multilinear objective, so that \cref{thm:uniform} bounds the bias in terms of a centered norm of the payoff tables. In \cref{sec:variance}, we bound the variance of the sampled optimum by an Efron--Stein argument. We combine the two bounds in \cref{sec:estimator} and treat graphs in \cref{sec:bip}. To count queries, we assume that a table query specifies an $r$-element set $e$ and returns $g_e$; a variable is examined if it belongs to a queried set.

\subsection{Encoding Max-\texorpdfstring{$r$}{r}CSPs}\label{sec:encoding}

We first fix notation for Max-$r$CSPs and the quantities that control the bias and the fluctuations of the estimator. Fix an integer $r\ge2$. Let $\cA_1,\dots,\cA_n$ be nonempty finite alphabets, and for each $e\in\binom{[n]}r$ let $g_e\colon\prod_{i\in e}\cA_i\to\R$ be a payoff table; an absent constraint is the zero table. Every table is indexed by $r$ distinct variables. For $S\subseteq[n]$ and $\mathbf{p}\in[0,1]^n$, define
\[
\Opt_g(S):=\max_{x\in\prod_{i\in S}\cA_i}\ \sum_{e\in\binom Sr}g_e(x_e),
\qquad
\Opt_{g,\mathbf{p}}:=\max_{x\in\prod_i\cA_i}\ \sum_{e\in\binom{[n]}r}\Bigl(\prod_{i\in e}p_i\Bigr)g_e(x_e),
\]
with $\Opt_g(S)=0$ when $\abs S<r$, and write $\Opt_g:=\Opt_g([n])$. Thus $\Opt_{g,\mathbf{p}}$ is the optimum of the mean objective, in which each constraint is weighted by the probability that all its variables are retained; for uniform retention it equals $p^r\Opt_g$. Let $\bar g_e$ be the average of $g_e$ over its table, put $g_e^\circ:=g_e-\bar g_e$, and define the \emph{centered energy}
\begin{equation}\label{eq:energy}
\cE(g)^2:=\sum_{e\in\binom{[n]}r}\ \sum_{a\in\prod_{i\in e}\cA_i}\bigl(g_e(a)-\bar g_e\bigr)^2.
\end{equation}
The hindsight gain is unchanged when each table is centered (see the proof of \cref{thm:csp-restriction}), so the centered energy will play the role of $\fro A$. When $0\le g_e\le w_e$ for nonnegative weights $w_e$, write $Q(w):=\sum_e w_e^2$; these range bounds will control the fluctuations in \cref{sec:variance}. If every alphabet has size at most $k$, then
\begin{equation}\label{eq:energy-range}
\cE(g)^2\le\frac{k^r}{4}\,Q(w).
\end{equation}
Indeed, a uniformly random table entry in $[0,w_e]$ has variance at most $w_e^2/4$, and the table has at most $k^r$ entries. For MaxCut, $r=2$, $\cA_i=\{0,1\}$, and $g_{\{i,j\}}(a,b)=\one[a\ne b]$ on graph edges. Each nonzero table has mean $1/2$ and centered energy $1$, so $\cE(g)^2=Q(w)=e(G)$ when $w$ is the adjacency matrix of $G$.

With this notation, \cref{thm:noise,thm:uniform} give the following bounds on the hindsight gain of a Max-$r$CSP.

\begin{theorem}[Hindsight gain for Max-$r$CSPs]\label{thm:csp-restriction}
For each fixed $r\ge2$ there is a constant $C_r$, independent of the alphabets, such that the following holds. If $S$ contains each variable $i$ independently with probability $p_i$, then
\begin{equation}\label{eq:csp-nonuniform}
0\le\E\Opt_g(S)-\Opt_{g,\mathbf{p}}
\le C_r\,\cE(g)\Bigl(\sum_i p_i\Bigr)^{(r-1)/2}
\sqrt{\max_i p_i(1-p_i)}.
\end{equation}
For uniform retention with probability $p\in(0,1]$,
\begin{equation}\label{eq:csp-uniform}
0\le\E\Opt_g(S_p)-p^r\Opt_g
\le C_r p^{r-1/2}\sqrt{1-p}\,n^{(r-1)/2}\cE(g).
\end{equation}
\end{theorem}

\begin{proof}
We first remove the table averages, which do not change the gain, and then encode the centered instance in our model.

\emph{Centering.} For every $S$, $\Opt_g(S)=\sum_{e\in\binom Sr}\bar g_e+\Opt_{g^\circ}(S)$. The corresponding identity for $\Opt_{g,\mathbf{p}}$ and independence of the variable indicators show that the hindsight gain is unchanged when every table is replaced by $g_e^\circ$. We may therefore assume that all tables have mean zero.

\emph{One-hot encoding.} Relabel each alphabet as $\cA_i=[k_i]$, take $D=\max_i k_i$, and let $K_i:=\{\mathbf{e}_a:a\in[k_i]\}\subseteq\R^D$ consist of the corresponding coordinate unit vectors. For $e=\{i_1<\cdots<i_r\}$, define
\[
A_e(x_e):=\sum_{a_j\in[k_{i_j}],\ j\in[r]}
g_e(a_1,\ldots,a_r)\prod_{j=1}^r x_{i_j,a_j}.
\]
The coefficients outside the alphabet coordinates are zero. On unit vectors representing an assignment $\sigma$, this interaction equals $g_e(\sigma_e)$. Thus for every $\mathbf{p}\in[0,1]^n$,
\[
f_{A,\mathbf{p}}(\mathbf{e}_{\sigma_1},\ldots,\mathbf{e}_{\sigma_n})
=\sum_{e\in\binom{[n]}r}\Bigl(\prod_{i\in e}p_i\Bigr)g_e(\sigma_e),
\qquad
\fro A^2=\cE(g)^2.
\]
The last equality uses the centering performed above. By the first identity, the vector program has the same weighted and restricted optima as the Max-$r$CSP, and \cref{thm:noise,thm:uniform} give the two bounds.
\end{proof}

\subsection{Sampling-Sensitive Variance Bounds}\label{sec:variance}

\Cref{thm:csp-restriction} controls the bias of the estimator, and it remains to control its fluctuations. Adding or removing one variable affects only the interactions whose other variables were all sampled, so we charge the variable only for these interactions rather than for its full degree. The resulting variance bound is on the same scale as the squared bias. Since we apply it to several quantities, we state it for any set function with this bounded-difference property.

\begin{lemma}[Sampling-sensitive variance]\label{lem:variance}
Let $w_e\ge0$ for $e\in\binom{[n]}r$, let $Q:=\sum_e w_e^2$, and suppose $f\colon2^{[n]}\to\R$ satisfies
\begin{equation}\label{eq:local-lip}
\abs{f(U\cup\{i\})-f(U)}
\le\sum_{\substack{e\ni i\\e\setminus\{i\}\subseteq U}}w_e
\qquad(U\subseteq[n],\ i\notin U).
\end{equation}
Then, for every $p\in[0,1]$,
\begin{equation}\label{eq:variance-r}
\Var f(S_p)\le r(1-p)p^r(1+pn)^{r-1}Q.
\end{equation}
For $r=2$ the sharper bound
\begin{equation}\label{eq:variance-two}
\Var f(S_p)\le2p^2(1-p)^2Q+2p^3(1-p)(n-1)Q
\end{equation}
also holds.
\end{lemma}

\begin{proof}
The statement is immediate if $n<r$, so assume $n\ge r$. Write $\xi_i:=\one[i\in S_p]$ and
\[
D_i:=\sum_{e\ni i}w_e\prod_{j\in e\setminus\{i\}}\xi_j.
\]
Thus $D_i$ is the total weight of the interactions at $i$ whose other variables are all sampled. By~\eqref{eq:local-lip}, replacing $\xi_i$ by an independent copy $\xi_i'$ changes $f(S_p)$ by at most $\abs{\xi_i-\xi_i'}D_i$. Since $D_i$ is independent of these two indicators and $\E(\xi_i-\xi_i')^2=2p(1-p)$, the Efron--Stein inequality~\cite{BLM13} gives
\begin{align*}
\Var f(S_p)
&\le p(1-p)\sum_i\E D_i^2\\
&=(1-p)\sum_{s=1}^r s p^{2r-s}
\sum_{\substack{e,e'\in\binom{[n]}r\\|e\cap e'|=s}}w_ew_{e'}.
\end{align*}
The equality follows by expanding $\E D_i^2$ over pairs $e,e'\ni i$: a pair with $|e\cap e'|=s$ appears for each of the $s$ choices of $i\in e\cap e'$, and the other $2r-s-1$ variables of $e\cup e'$ are all sampled with probability $p^{2r-s-1}$. The last sum is over ordered pairs. For each fixed $e$, there are $\binom rs\binom{n-r}{r-s}$ possible $e'$ with $|e\cap e'|=s$. The inequality $2w_ew_{e'}\le w_e^2+w_{e'}^2$ therefore yields
\begin{equation}\label{eq:variance-overlap}
\Var f(S_p)
\le(1-p)Q\sum_{s=1}^r s\binom rs\binom{n-r}{r-s}p^{2r-s}.
\end{equation}
Putting $j=r-s$, using $(r-j)\binom rj=r\binom{r-1}j$ and $\binom{n-r}j\le n^j$, and summing the binomial expansion proves~\eqref{eq:variance-r}. For $r=2$, the right-hand side of~\eqref{eq:variance-overlap} is
$2p^2(1-p)[1+(n-2)p]Q$, which equals~\eqref{eq:variance-two}.
\end{proof}

Before applying the lemma, we check condition~\eqref{eq:local-lip} for the functions we need. It holds for $f=\Opt_g$ whenever $0\le g_e\le w_e$: an optimal assignment on $U$ extends without loss, while restricting an optimal assignment on $U\cup\{i\}$ loses at most the sum of the newly present payoffs. It also holds when $|g_e|\le w_e$, by using arbitrary extensions and restrictions for the two inequalities. For graphs it holds, with $r=2$ and $w$ the adjacency matrix, for the number of edges, the maximum cut, and the distance to bipartiteness. For the last quantity, deleting a vertex cannot increase the distance, and deleting all its incident edges together with an optimal deletion set for the remaining graph makes the enlarged graph bipartite.

\subsection{The Estimator}\label{sec:estimator}

We now combine the bias and variance bounds into an accuracy guarantee. For $1\le q\le n$, we use the following estimator.

\begin{center}
\begin{minipage}{0.9\textwidth}
\textbf{Estimator $\mathsf{Est}_q$.} Put $p:=q/n$ and sample $S=S_p$. If $\abs S>4q$, output $0$. Otherwise read the tables $g_e$ with $e\in\binom Sr$ and output $\widehat{\Opt}:=p^{-r}\Opt_g(S)$.
\end{minipage}
\end{center}

The sample is chosen before any table is read, so the estimator is nonadaptive, and the cutoff at $4q$ ensures that it reads $O_r(q^r)$ tables.

\begin{theorem}[Energy-sensitive estimator]\label{thm:csp-estimator}
For every fixed $r\ge2$ there is a constant $C_r$ such that the following holds. Let $0\le g_e\le w_e$ and $1\le q\le n$. With probability at least $5/6$, the estimator $\mathsf{Est}_q$ samples at most $4q$ variables and satisfies
\[
\bigl|\widehat{\Opt}-\Opt_g\bigr|
\le C_r\,\frac{n^{r/2}}{\sqrt q}\Bigl(\cE(g)+\sqrt{Q(w)}\Bigr).
\]
\end{theorem}

\begin{proof}
Let $p=q/n$ and $Z:=\Opt_g(S_p)$. We bound the bias and the fluctuations of $p^{-r}Z$ separately and then account for the cutoff. By~\eqref{eq:csp-uniform}, the bias satisfies
\[
0\le p^{-r}\E Z-\Opt_g
\le C_r p^{-1/2}n^{(r-1)/2}\cE(g)
=C_r\frac{n^{r/2}}{\sqrt q}\cE(g).
\]
For the fluctuations, since $\Opt_g$ satisfies~\eqref{eq:local-lip}, \cref{lem:variance} and $q\ge1$ give
\[
\Var(p^{-r}Z)
\le r p^{-r}(1+q)^{r-1}Q(w)
\le r2^{r-1}\frac{n^r}{q}Q(w).
\]
Chebyshev's inequality bounds $\abs{p^{-r}Z-p^{-r}\E Z}$ by
$\sqrt{12r2^{r-1}}\,n^{r/2}\sqrt{Q(w)/q}$ with probability at least $11/12$. It remains to account for the cutoff. The sample size $\abs{S_p}$ is a sum of independent indicators with mean $q\ge1$, so the Chernoff bound gives
$\Prob(\abs{S_p}>4q)\le(e^3/4^4)^q<1/12$. On the event $\abs{S_p}\le4q$ the estimator outputs $p^{-r}Z$, so a union bound proves the claim, with an adjusted $C_r$.
\end{proof}

\begin{corollary}[Bounded alphabets]\label{cor:csp-k}
For every fixed $r\ge2$ there is a constant $C_r$ such that the following holds for every $\eps\in(0,1)$. If all alphabets have size at most $k$ and all payoffs lie in $[0,1]$, take
\[
q:=\min\{n,C_r k^r\eps^{-2}\}.
\]
With probability at least $5/6$, $\mathsf{Est}_q$ samples at most $4q$ variables and satisfies $\abs{\widehat{\Opt}-\Opt_g}\le\eps n^r$. It reads $O_r(k^{r^2}\eps^{-2r})$ payoff tables.
\end{corollary}

\begin{proof}
If $q=n$, the estimator is exact. Otherwise take $w_e:=\one[g_e\not\equiv0]$. Then $Q(w)\le\binom nr\le n^r$ and, by~\eqref{eq:energy-range}, $\cE(g)\le\frac12k^{r/2}\sqrt{Q(w)}$. The error in \cref{thm:csp-estimator} is at most $C_r'k^{r/2}n^r/\sqrt q$, which is at most $\eps n^r$ after increasing $C_r$.
\end{proof}

This proves \cref{thm:csp-intro}. The $O(\eps^{-2})$ dependence on the accuracy cannot be improved for any fixed $r$ and alphabet size at least two, since \cref{cor:csp-vertices} gives a matching lower bound on the number of examined variables. When the tables are sparse, \cref{thm:csp-estimator} can give a better alphabet dependence than \cref{cor:csp-k}. For example, in a pairwise Unique Games instance over $k$ labels, each nonzero table is a permutation matrix with squared centered energy $k-1$, so $O(k\eps^{-2})$ sampled variables suffice, compared with $O(k^2\eps^{-2})$ from \cref{cor:csp-k}.

\begin{remark}[Constraints of smaller arity]\label{rem:smaller-arity}
The same $O_{r,k}(\eps^{-2})$ sample bound holds when there is a payoff table with values in $[0,1]$ on each nonempty subset of at most $r$ variables. If $q=n$, the estimator is exact. Otherwise, apply \cref{cor:csp-k} to the $r$-variable tables. The remaining tables change the original optimum by at most $C_r n^{r-1}$ and, on the event $\abs S\le4q$, change the sampled optimum by at most $C_r q^{r-1}$. Their total contribution to the rescaled error is at most
$C_r(n^{r-1}+p^{-r}q^{r-1})\le2C_r n^r/q$,
which is absorbed in the stated error after adjusting the sample-size constant.
\end{remark}

\subsection{MaxCut and the Distance to Bipartiteness}\label{sec:bip}

We now specialize to graphs. For MaxCut, the estimator reads the subgraph induced by the sample and rescales its maximum cut. Since $\beta(G)=e(G)-\mc(G)$, estimating the number of edges from the same sample also gives an estimate of the distance to bipartiteness.

\begin{corollary}[Graph estimator]\label{cor:bip-upper}
There is an absolute constant $C$ such that the following holds for every $\eps\in(0,1)$. Let $G$ be a graph on $n$ vertices, let $q:=\min\{n,C\eps^{-2}\}$ and $p:=q/n$, and sample $S=S_p$. If $\abs{S}>4q$, output $0$; otherwise query all pairs inside $S$ and output $\widehat{\mc}:=p^{-2}\mc(G[S])$ and $\widehat\beta:=p^{-2}\beta(G[S])$. Then, with probability at least $3/4$, both outputs are within $\pm\eps n^2$ of $\mc(G)$ and $\beta(G)$, respectively. The algorithm is nonadaptive, examines at most $4q$ vertices, and makes fewer than $8q^2$ queries.
\end{corollary}

\begin{proof}
If $q=n$, both outputs are exact, so assume $q<n$. Let $m:=e(G)\le n^2/2$, and let $w$ be the adjacency matrix of $G$, so that $\cE(g)^2=Q(w)=m$ for MaxCut. By \cref{thm:csp-estimator} with $r=2$, with probability at least $5/6$ we have $\abs S\le4q$ and $\abs{\widehat{\mc}-\mc(G)}\le C'n^2/\sqrt q$ for an absolute constant $C'$. To estimate $\beta(G)=e(G)-\mc(G)$, we also need the number of edges. It satisfies~\eqref{eq:local-lip} as well, so $p^{-2}e(G[S_p])$, whose mean is $m$, has variance at most $2p^{-2}(1+q)m\le2n^4/q$ by \cref{lem:variance}; by Chebyshev's inequality, it is within $\sqrt{24}\,n^2/\sqrt q$ of $m$ with probability at least $11/12$. On both events, $\widehat\beta=p^{-2}e(G[S])-\widehat{\mc}$ and $\beta(G)=m-\mc(G)$ differ by at most $(C'+\sqrt{24})\,n^2/\sqrt q$. Taking $C\ge(C'+\sqrt{24})^2$ proves the accuracy claims with probability at least $3/4$. Finally, the algorithm makes at most $\binom{\abs S}2<8q^2$ queries.
\end{proof}

\Cref{cor:bip-upper} proves \cref{thm:bip-intro}\ref{it:bip-upper}. Apart from sampling $S$, the running time is $2^{O(\eps^{-2})}$, by exhaustive search over cuts. The matching lower bounds on examined vertices and nonadaptive queries are proved in \cref{sec:lower}.

\section{Application II: Max-Cut under Vertex Percolation}\label{sec:zhao}

Vertex percolation at rate $p$ keeps each vertex of a graph independently with probability $p$. Zhao~\cite{OWR19} asked whether, for $p=\frac12$, it preserves the max-cut fraction of every sequence of graphs whose ratio of edges to vertices tends to infinity. In this application, we use our bound on the hindsight gain for MaxCut to answer this question for every retention rate.

Let $G$ be a graph with $n$ vertices and $m$ edges, put $\rho:=m/n$, and let $H:=G[S_p]$. We compare the max-cut fraction $\mc(H)/e(H)$ of the sample with $\mc(G)/e(G)$. We will show that, when $p\rho\ge24/\delta$, the two fractions differ by $O\bigl(\sqrt{(1-p)/(\delta p\rho)}\bigr)$ with probability at least $1-\delta$ (\cref{thm:zhao}). Letting $\delta$ tend to zero slowly then answers Zhao's question (\cref{cor:zhao}), and the complete graph shows that the condition $p\rho\to\infty$ cannot be weakened (\cref{prop:zhao-sharp}).

The max-cut fraction of $H$ is a ratio of two random quantities, so we control the numerator and the denominator separately. The case $r=2$ of \cref{thm:csp-restriction} bounds the mean of $\mc(H)$, and the graph bound~\eqref{eq:variance-two} controls the fluctuations of both $\mc(H)$ and $e(H)$. Along the way we also obtain a multiplicative estimate for the rescaled max-cut, which we use in \cref{rem:coreset}.

\begin{theorem}[Quantitative vertex percolation]\label{thm:zhao}
There is an absolute constant $C$ such that the following holds. Let $G$ be a graph with $n$ vertices and $m\ge1$ edges, let $\rho:=m/n$, and let $p\in(0,1]$ and $\delta\in(0,1)$ satisfy $p\rho\ge24/\delta$. Then, with probability at least $1-\delta$, the graph $H:=G[S_p]$ satisfies $e(H)\ge\frac12p^2m>0$,
\[
\Bigl|\frac{\mc(H)}{e(H)}-\frac{\mc(G)}{e(G)}\Bigr|\le C\sqrt{\frac{1-p}{\delta\,p\rho}},
\qquad\text{and}\qquad
\Bigl|\frac{\mc(H)}{p^2\mc(G)}-1\Bigr|\le C\sqrt{\frac{1-p}{\delta\,p\rho}}.
\]
\end{theorem}

\begin{proof}
Let $Z_1:=\mc(H)$ and $Z_2:=e(H)$. We first compare their means with $p^2\mc(G)$ and $p^2m$, then show that both are concentrated, and finally bound the two ratios.

\emph{Means.} Each edge survives with probability $p^2$, so $\E Z_2=p^2m$. For the numerator, \eqref{eq:csp-uniform} with $\cE(g)^2=m$ gives
\begin{equation}\label{eq:zhao-bias}
0\le\E Z_1-p^2\mc(G)\le C_2p^{3/2}\sqrt{1-p}\,\sqrt{mn},
\end{equation}
where $C_2$ is the constant in~\eqref{eq:csp-uniform} for $r=2$.

\emph{Fluctuations.} Both $Z_1$ and $Z_2$ satisfy the hypothesis of \cref{lem:variance} with $w$ the adjacency matrix of $G$, so $Q(w)=m$ and
\[
\Var(Z_k)\le\sigma^2:=2p^2(1-p)^2m+2p^3(1-p)mn,\qquad k=1,2 .
\]
By Chebyshev's inequality and a union bound, with probability at least $1-\delta$ we have $\abs{Z_k-\E Z_k}\le u:=\sqrt{2\sigma^2/\delta}$ for $k=1,2$. Put $\eta:=u/(p^2m)$, the deviation relative to $\E Z_2$, so that
\[
\eta^2=\frac{4(1-p)}{\delta}\Bigl(\frac{1-p}{p^2m}+\frac{1}{p\rho}\Bigr).
\]
To control the ratio, we first show that this relative fluctuation is small enough to keep the denominator away from zero.
Since $m\le\binom n2$, we have $n\ge2\rho+1$, and hence $p^2m=p^2\rho n\ge2(p\rho)^2\ge2p\rho$, as $p\rho\ge1$. Therefore $\eta^2\le6(1-p)/(\delta p\rho)\le\frac14$.

\emph{Conclusion.} On the event above, $Z_2\ge(1-\eta)p^2m\ge\frac12p^2m$. With $a:=\mc(G)/m\le1$ we have $Z_1-aZ_2=(Z_1-p^2\mc(G))-a(Z_2-p^2m)$, so by~\eqref{eq:zhao-bias},
\[
\Bigl|\frac{Z_1}{Z_2}-a\Bigr|\le\frac{2u+C_2p^{3/2}\sqrt{1-p}\sqrt{mn}}{\frac12p^2m}=4\eta+2C_2\sqrt{\frac{1-p}{p\rho}} .
\]
Similarly, since $\mc(G)\ge m/2$, we have $\abs{Z_1/(p^2\mc(G))-1}\le2\eta+2C_2\sqrt{(1-p)/(p\rho)}$. Both bounds are at most $(4\sqrt6+2C_2)\sqrt{(1-p)/(\delta p\rho)}$.
\end{proof}

We now pass to sequences of graphs.

\begin{corollary}[Zhao's question]\label{cor:zhao}
Let $(G_N)$ be graphs with $e(G_N)\ge1$, let $\rho_N:=e(G_N)/v(G_N)$, and let $p_N\in(0,1]$ satisfy $p_N\rho_N\to\infty$. Then, with probability tending to one, $H_N:=G_N[S_{p_N}]$ has an edge and
\[
\Bigl|\frac{\mc(H_N)}{e(H_N)}-\frac{\mc(G_N)}{e(G_N)}\Bigr|\le C(p_N\rho_N)^{-1/4},
\]
where $C$ is the constant of \cref{thm:zhao}. In particular, for $p_N=\frac12$ and $\rho_N\to\infty$, the max-cut fraction is preserved with probability tending to one.
\end{corollary}

\begin{proof}
To make both the failure probability and the error bound tend to zero, apply \cref{thm:zhao} with $\delta_N:=24(p_N\rho_N)^{-1/2}$. This tends to zero, and it lies in $(0,1)$ and satisfies $p_N\rho_N\ge24/\delta_N$ as soon as $p_N\rho_N>576$. The error bound becomes $C\bigl(24(p_N\rho_N)^{1/2}\bigr)^{-1/2}\le C(p_N\rho_N)^{-1/4}$.
\end{proof}

Conversely, the density condition cannot be weakened, as the complete graph shows.

\begin{proposition}[Sharpness of the density condition]\label{prop:zhao-sharp}
For every $c>0$ there are graphs $G_N$ with $\rho_N\to\infty$ and probabilities $p_N$ with $p_N\rho_N=c$ such that, with probability bounded away from zero, $G_N[S_{p_N}]$ has at most one edge, whereas $\mc(G_N)/e(G_N)\to\frac12$. In particular, the max-cut fraction of $G_N$ is not preserved with probability tending to one.
\end{proposition}

\begin{proof}
Let $G_N=K_N$, so that $\rho_N=(N-1)/2$, and let $p_N:=2c/(N-1)$ for $N>2c+1$. Then $\abs{S_{p_N}}$ converges in distribution to a Poisson variable with mean $2c$, so $\Prob(\abs{S_{p_N}}\le2)\to e^{-2c}(1+2c+2c^2)>0$. On this event $G_N[S_{p_N}]$ either has no edge or consists of a single edge, whose max-cut fraction is $1$, whereas $\mc(K_N)/e(K_N)\to\frac12$.
\end{proof}

\Cref{cor:zhao,prop:zhao-sharp} together prove \cref{thm:zhao-intro}.

\begin{remark}[Uniform core-sets for MaxCut]\label{rem:coreset}
The second estimate of \cref{thm:zhao} is a multiplicative guarantee for the rescaled sampled max-cut. For a graph of average degree $\bar d=2\rho$, sample each vertex with probability $p=\min\{1,C\eps^{-2}/\bar d\}$ for a suitable absolute constant $C$. Taking $\delta=1/3$ in \cref{thm:zhao}, we obtain $\abs{p^{-2}\mc(G[S_p])/\mc(G)-1}\le\eps$ with probability at least $2/3$, and the expected number of sampled vertices is $O(\eps^{-2}n/\bar d)$. For $\bar d=n^\alpha$ this is $O(\eps^{-2}n^{1-\alpha})$, the same polynomial dependence on $n$ as the core-sets of Bhaskara, Daruki, and Venkatasubramanian~\cite{BDV18}, which is optimal under the exponential time hypothesis. Their core-sets use degree-biased sampling and succeed with high probability, whereas our guarantee uses uniform sampling and holds with constant probability.
\end{remark}

\begin{remark}[One-half percolation]\label{rem:half}
For $p=\frac12$, the case of Zhao's question, the bounded differences inequality~\cite{McD89} gives exponential tails in place of Chebyshev's inequality, and hence a bound that holds with high probability. Let $m=e(G)$ and $n=v(G)$. Changing the membership of a single vertex $v$ changes $\mc(H)$ and $e(H)$ by at most $\deg_G(v)$, and $\sum_v\deg_G(v)^2\le2m(n-1)<2mn$. The bounded differences inequality with deviation $u:=\sqrt{3mn\log\rho}$, together with a union bound, therefore shows that with probability at least $1-4\rho^{-3}$ both quantities are within $u$ of their means. If $\rho$ is so large that $u/m=\sqrt{3\log\rho/\rho}\le\frac18$, then $e(H)\ge\frac m4-u\ge\frac m8$. Combining these bounds with~\eqref{eq:zhao-bias} at $p=\frac12$, exactly as in the proof of \cref{thm:zhao}, gives
\[
\Bigl|\frac{\mc(H)}{e(H)}-\frac{\mc(G)}{e(G)}\Bigr|\le\frac8m\Bigl(2u+\frac{C_2}{4}\sqrt{mn}\Bigr)=O\Bigl(\sqrt{\frac{\log\rho}{\rho}}\Bigr).
\]
\end{remark}

\section{Application III: Sampling Quantum Ground-State Energies}\label{sec:quantum}

A local quantum Hamiltonian describes a many-particle system through interactions involving a bounded number of particles. Its ground-state energy is its minimum eigenvalue: the lowest energy attainable by any state, including entangled states. In this application, we use our uniform-retention theorem (\cref{thm:uniform}) to estimate this energy from the interactions among a small random sample of particles.

We begin with interactions indexed by $r$-element sets, and allow terms of smaller arity at the end of the proof:
\[
H=\sum_{e\in\binom{[n]}r}h_e,
\qquad \opn{h_e}\le1,
\]
where each $h_e$ is Hermitian and acts on the particles in $e$, each of local dimension $d$. Identity operators on the remaining particles are implicit, and zero interactions are allowed. For $U\subseteq[n]$, let $H_U:=\sum_{e\in\binom Ur}h_e$ and $E_0(H_U):=\lambda_{\min}(H_U)$.

Given a uniformly random set $S$ of $q$ particles, we estimate $E_0(H)$ by the rescaled sampled energy $(n/q)^rE_0(H_S)$. For fixed $r\ge2$ and $d$, and $\eps\in(0,1)$, we will show that $O_{r,d}(\eps^{-3})$ sampled particles suffice for additive error $\eps n^r$ with probability at least $0.99$, taking the whole instance if the required sample size exceeds $n$. This is the guarantee of \cref{thm:quantum-intro}.

Our proof proceeds in two steps. We first restrict to pure product states, which are tensor products of single-particle unit vectors. Their energy is multilinear in the corresponding rank-one density matrices, allowing us to obtain an $O_{r,d}(\eps^{-2})$ sample bound from our uniform-retention theorem. We then use a product-state approximation theorem to pass to the full ground-state energy. To carry out the first step, we define the minimum product-state energy by
\begin{equation}\label{eq:product-energy}
P(H_U):=\min_{\rho_i\in\mathsf P_d,\ i\in U}
\sum_{e\in\binom Ur}\tr\left(h_e\bigotimes_{i\in e}\rho_i\right),
\qquad
\mathsf P_d:=\{\psi\psi^*: \psi\in\mathbb C^d,\ \norm\psi_2=1\}.
\end{equation}
The factors in each tensor product are ordered by their particle labels. Since all interactions here are indexed by $r$-element sets, both energies are zero when $\abs U<r$. Restricting the state space to product states gives $E_0(H_U)\le P(H_U)$.

\subsection{Sampling Product-State Energies}\label{sec:quantum-product}

We first express the product-state objective in real coordinates. Fix an orthonormal basis $F_1,\ldots,F_{d^2}$ of the real vector space of Hermitian $d\times d$ matrices, with inner product $\ip XY=\tr(XY)$. We represent $\rho_i\in\mathsf P_d$ by the real vector $x_i=(\tr(\rho_iF_a))_{a=1}^{d^2}$. These vectors form a compact set $K_i\subseteq\R^{d^2}$, and
\[
\norm{x_i}_2^2=\tr(\rho_i^2)=1.
\]
Thus the feasible sets fit our model with $D=d^2$; convexity is not required. To turn energy minimization into maximization, we negate the interactions: for $e=\{i_1<\cdots<i_r\}$, define
\[
(A_e)_{a_1,\ldots,a_r}:=
-\tr\bigl(h_e(F_{a_1}\otimes\cdots\otimes F_{a_r})\bigr).
\]
Then $A_e(x_e)=-\tr(h_e\bigotimes_{i\in e}\rho_i)$, so $\Val_A(U)=-P(H_U)$. The coefficient norm needed for our sampling bound satisfies
\begin{equation}\label{eq:quantum-frobenius}
\fro A^2=\sum_{e\in\binom{[n]}r}\fro{A_e}^2
=\sum_{e\in\binom{[n]}r}\fro{h_e}^2
\le d^r\binom nr
\le d^r n^r.
\end{equation}
Here $\fro{h_e}$ is the Frobenius norm of its local $d^r\times d^r$ matrix, without the implicit identity factors, so $\fro{h_e}^2\le d^r\opn{h_e}^2\le d^r$. The middle equality follows because the tensor products of the $F_a$ form an orthonormal basis for the Hermitian operators on $r$ particles.

\begin{lemma}[Sampling product-state energies]\label{lem:quantum-product}
For every fixed $r\ge2$ there is a constant $C_r$ such that, for every integer $1\le q\le n$ and a uniformly random $q$-element set $S\subseteq[n]$, with probability at least $0.99$,
\begin{equation}\label{eq:product-sampling}
\left|\left(\frac nq\right)^rP(H_S)-P(H)\right|
\le C_r d^{r/2}\,\frac{n^r}{\sqrt q}.
\end{equation}
In particular, $O_r(d^r\eps^{-2})$ sampled particles suffice for additive error $\eps n^r$, taking the whole instance if this number exceeds $n$.
\end{lemma}

\begin{proof}
The case $q=n$ is immediate. Put $p:=q/n$ and first use a Bernoulli sample $S_p$, for which our uniform-retention theorem controls the bias directly. By \cref{thm:uniform} and~\eqref{eq:quantum-frobenius},
\begin{equation}\label{eq:product-bias}
0\le P(H)-p^{-r}\E P(H_{S_p})
\le C_r p^{-1/2}n^{(r-1)/2}\fro A
\le C_r d^{r/2}\frac{n^r}{\sqrt q}.
\end{equation}

To control fluctuations, we bound how much adding one particle can change the optimum. For every product state, each summand in~\eqref{eq:product-energy} lies in $[-1,1]$. Restricting an optimal product state from $U\cup\{i\}$ to $U$, and extending an optimal product state from $U$ by an arbitrary state at $i$, give
\begin{equation}\label{eq:quantum-add-one}
\abs{P(H_{U\cup\{i\}})-P(H_U)}\le\binom{\abs U}{r-1}.
\end{equation}
This is~\eqref{eq:local-lip} with $w_e=1$, so we can apply \cref{lem:variance} with $Q=\binom nr\le n^r$. Since $q=pn\ge1$,
\[
\Var\bigl(p^{-r}P(H_{S_p})\bigr)
\le r p^{-r}(1+q)^{r-1}\binom nr
\le C_r\frac{n^{2r}}q.
\]
Combining this variance bound with Chebyshev's inequality and~\eqref{eq:product-bias}, we obtain
\begin{equation}\label{eq:product-bernoulli}
\left|p^{-r}P(H_{S_p})-P(H)\right|
\le C_r d^{r/2}\frac{n^r}{\sqrt q}
\end{equation}
with probability at least $1-1/200$, after adjusting $C_r$.

It remains to pass from Bernoulli sampling to a fixed-size sample. We couple the two samples using a uniformly random permutation of $[n]$ and an independent $N\sim\operatorname{Bin}(n,p)$. Let $U$ and $S$ be its first $N$ and first $q$ elements, respectively. Then $U$ has the law of $S_p$ and $S$ is uniform among the $q$-element subsets. With probability at least $1-1/200$, $\abs{N-q}\le\sqrt{200q}$. Because the sets are nested, we can pass from one to the other by adding or removing one particle at a time. On this event,~\eqref{eq:quantum-add-one} therefore gives
\[
\abs{P(H_U)-P(H_S)}
\le\abs{N-q}\binom{\max\{N,q\}}{r-1}
\le C_r q^{r-1/2}.
\]
Multiplying by $p^{-r}$ and combining with~\eqref{eq:product-bernoulli} by a union bound proves the lemma.
\end{proof}

\subsection{From Product States to the Ground-State Energy}\label{sec:quantum-ground}

To transfer our sampling estimate to the ground-state energy, we need to bound the error from restricting to product states. We use Bergamaschi's product-state approximation theorem~\cite[Theorem~3]{Bergamaschi23}, which builds on Brand\~ao and Harrow~\cite{BH13}. For a Hamiltonian $H'$ on $s$ particles of local dimension $d$, with $m$ nonzero interactions indexed by $r$-element subsets and of norm at most one,
\begin{equation}\label{eq:quantum-product-gap}
0\le P(H')-E_0(H')
\le B_{r,d}s^{(r-1)/3}m^{2/3}
\le B_{r,d}s^{r-1/3},
\end{equation}
where the last inequality uses $m\le\binom sr$. No regularity assumption on the interaction hypergraph is required. This is important for us, since we apply the bound both to $H$, which is arbitrary, and to the sampled Hamiltonian $H_S$.

We now restate the ground-state guarantee, allowing at most one interaction of norm at most one on each nonempty subset of at most $r$ particles, with $H_S$ retaining all terms supported in $S$.

\ThmQuantum*

\begin{proof}[Proof of \cref{thm:quantum-intro}]
If $q=n$, the estimate is exact. Otherwise, we compare the normalized energies $E_0(H_S)/q^r$ and $E_0(H)/n^r$ through the corresponding product-state energies. By~\eqref{eq:quantum-product-gap} and $q\le n$, the normalized gaps $q^{-r}(P(H_S)-E_0(H_S))$ and $n^{-r}(P(H)-E_0(H))$ both lie in $[0,B_{r,d}q^{-1/3}]$. Their difference therefore has absolute value at most $B_{r,d}q^{-1/3}$. Combining this with \cref{lem:quantum-product}, we obtain with probability at least $0.99$ that
\begin{align*}
\left|\frac{E_0(H_S)}{q^r}-\frac{E_0(H)}{n^r}\right|
&\le \left|\frac{P(H_S)}{q^r}-\frac{P(H)}{n^r}\right|
 +B_{r,d}q^{-1/3}\\
&\le C_r d^{r/2}q^{-1/2}+B_{r,d}q^{-1/3}
\le C'_{r,d}q^{-1/3},
\end{align*}
since $q\ge1$. Multiplying by $n^r$ and choosing the constant in the sample size at least $(C'_{r,d})^3$ proves the assertion for interactions indexed by $r$-element sets.

If $H$ also contains terms on fewer than $r$ particles, we extend the sums defining $H_U$ and $P(H_U)$ to all $e\subseteq U$ with $1\le\abs e\le r$. We treat these additional terms as a perturbation: write $H=H^{(r)}+R$, where $H^{(r)}$ consists of its $r$-particle terms. With at most one term of norm at most one on each nonempty subset, we have
\[
\opn R\le\sum_{j=1}^{r-1}\binom nj\le C_r n^{r-1},
\qquad
\opn{R_S}\le C_r q^{r-1}.
\]
The variational characterization of the minimum eigenvalue bounds the effect of this perturbation:
$\abs{E_0(H)-E_0(H^{(r)})}\le\opn R$, and similarly for the sample. Hence the additional error in the rescaled estimate is at most
\[
C_r\left(n^{r-1}+\left(\frac nq\right)^r q^{r-1}\right)
\le 2C_r\frac{n^r}q.
\]
We can absorb this error into $C'_{r,d}n^r q^{-1/3}$ by adjusting the constant, which proves the stated $r$-local result.
\end{proof}

The same perturbation argument applies to product-state minima, so their sampling estimate also permits interactions of smaller arity. Our product-state estimate needs $O_{r,d}(\eps^{-2})$ sampled particles, and our ground-state estimate needs $O_{r,d}(\eps^{-3})$; both are additive guarantees on the scale $n^r$. For fixed $r,d\ge2$ and sufficiently large $n$, the $\eps^{-2}$ dependence of the product-state sample bound is optimal by \cref{cor:quantum-particles}, even for diagonal qubit Hamiltonians. That corollary also gives an $\Omega_r(\eps^{-2})$ lower bound for ground-state energy. The exponent $3$ in our ground-state upper bound is inherited from~\eqref{eq:quantum-product-gap}, and we do not claim it is optimal.

To implement the ground-state estimator, we query the interactions within the sampled set. If a query returns the classical description of a local interaction, this uses $O_{r,d}(\eps^{-3r})$ nonadaptive queries. Computing the minimum eigenvalue of the sampled Hamiltonian may take time exponential in $q$.

\section{Conclusion}\label{sec:open}

Random restrictions change both the available variables and the solution chosen after they are observed. In this paper, we isolate the value of this second effect, the hindsight gain, by comparing optimal values through a support direction. For homogeneous multilinear objectives, this gain is controlled by the Frobenius norm of the interaction coefficients and second moments of the weights. Under uniform retention, we iterate the comparison across scales and give an inverse-square-root dependence on the expected sample size. These bounds yield optimal accuracy dependence for sampling dense Max-$r$CSPs, sharp graph estimation bounds, preservation of max-cut fractions under vertex percolation, and a cubic sampling bound for quantum ground-state energies.

We conclude with several open questions.

\paragraph{Adaptive queries.}
The adaptive query complexity of estimating $\mc(G)$ or $\beta(G)$ within $\pm\eps n^2$ lies between $\Omega(\eps^{-2})$ and $O(\eps^{-4})$ (\cref{sec:lower}). An adaptive upper bound of $o(\eps^{-4})$ would mean that the optimum can be estimated without reading the full subgraph induced by the examined vertices, since every algorithm must examine $\Omega(\eps^{-2})$ of them. Whether such adaptive algorithms exist remains open.

\paragraph{Nonuniform retention.}
The refined estimate (\cref{thm:refined}) allows arbitrary inclusion probabilities and keeps track of the coefficient mass incident to each variable, but our sharp uniform estimate needs a common thinning rate. We ask whether, for independent Bernoulli weights with means $p_i$, one always has
\[
\E\Val_A(S)-\Val_A(\mathbf{p})
\le C_r\fro A
\left(\sum_i p_i^2\right)^{(r-1)/2}
\sqrt{\max_i p_i(1-p_i)}.
\]
For equal probabilities this is the scale of \cref{thm:uniform}. A bound that also retains the coefficient masses incident to individual variables could guide the choice of probabilities under a sampling budget.

\paragraph{Arity, alphabet size, and entanglement.}
The $\eps^{-2}$ sample dependence for Max-$r$CSPs is optimal for any fixed arity and alphabet, but the joint dependence on these parameters remains to be determined. In particular, can optimal accuracy dependence coexist with the polynomial arity and polylogarithmic alphabet dependence of~\cite{BCCFM26}? For quantum ground-state energies, the $\eps^{-3}$ bound comes from the product-state approximation error. Whether the ground-state sampling exponent can match the exponent $2$ for product-state energies is open.

\clearpage
\appendix
\crefalias{section}{appendix}
\section{Lower Bounds}\label{sec:lower}

In this appendix we prove \cref{thm:bip-intro}\ref{it:bip-lower}. We first show that nonadaptive algorithms need $\Omega(\eps^{-4})$ queries (\cref{thm:lower}), by the usual two-distribution argument: we construct two input distributions whose values of $\beta$, and of $\mc$, are separated by about $4\eps n^2$ with high probability, while the answers to any fixed set of fewer than $c\eps^{-4}$ queries have nearly the same law under both. We then deduce, by a random relabeling argument, that every algorithm must examine $\Omega(\eps^{-2})$ vertices (\cref{cor:vertices}). Finally, we transfer this bound to Max-$r$CSPs and local Hamiltonians (\cref{cor:csp-vertices,cor:quantum-particles}).

\begin{theorem}[Nonadaptive queries]\label{thm:lower}
There are absolute constants $c,C>0$ such that for every $\eps\in(0,1/64]$ and every $n\ge C\eps^{-2}$ the following holds. Let $\mathsf{Alg}$ be a randomized nonadaptive algorithm that makes at most $T$ adjacency queries and, for every $n$-vertex graph $G$, outputs with probability at least $2/3$ a number within $\eps n^2$ of $\beta(G)$ (respectively, of $\mc(G)$). Then $T\ge c\,\eps^{-4}$.
\end{theorem}

\paragraph{The two input distributions.}
Let $\theta:=32\eps\le\frac12$. Let $\cD_0$ be the Erd\H{o}s--R\'enyi distribution $G(n,\frac12)$. To sample from $\cD_1$, draw independent uniform labels $\sigma_u\in\{\pm1\}$ for $u\in[n]$, and then include each pair $uv$ independently with probability
\[
\Prob(uv\in E\mid\sigma)=\tfrac12\bigl(1-\theta\,\sigma_u\sigma_v\bigr).
\]
Pairs with different labels are thus slightly more likely to be edges than pairs with equal labels; this typically lowers $\beta$ and raises $\mc$ by about $4\eps n^2$ (\cref{lem:separation}). Averaged over the labels, however, every pair is an edge with probability exactly $\frac12$, so a single query has the same distribution under $\cD_0$ and $\cD_1$; the hidden bipartition reveals itself only through correlations between the answers to queries that form cycles.

\begin{lemma}[Separation]\label{lem:separation}
For $n\ge4\cdot10^4$, with probability at least $0.99$ a graph drawn from $\cD_0$ satisfies
\[
\beta(G)\ge\tfrac18n^2-3n^{3/2}\qquad\text{and}\qquad\mc(G)\le\tfrac18n^2+n^{3/2},
\]
and with probability at least $0.99$ a graph drawn from $\cD_1$ satisfies
\[
\beta(G)\le\tfrac18n^2-4\eps n^2+2n^{3/2}\qquad\text{and}\qquad\mc(G)\ge\tfrac18n^2+4\eps n^2-2n^{3/2}.
\]
\end{lemma}

\begin{proof}
All deviation bounds below are applications of Hoeffding's inequality to binomial variables on at most $\binom n2$ trials, with deviation $n^{3/2}$; each fails with probability at most $e^{-4n}$.

Under $\cD_0$, $e(G)\ge\frac12\binom n2-n^{3/2}$. For a fixed cut $(X,X^c)$, the number of cut edges is binomial on $\abs{X}\abs{X^c}\le n^2/4$ trials with mean at most $n^2/8$. A union bound over the $2^n$ cuts shows that $\mc(G)\le\frac18n^2+n^{3/2}$ with probability at least $1-2^ne^{-4n}$. Hence $\beta(G)=e(G)-\mc(G)\ge\frac18n^2-\frac n4-2n^{3/2}\ge\frac18n^2-3n^{3/2}$.

Under $\cD_1$, let $R:=\sum_u\sigma_u$. Since $\E R^2=n$, Chebyshev's inequality gives $\abs{R}\le n^{3/4}$ with probability at least $1-n^{-1/2}$. On this event the numbers of equal-label and different-label pairs are
\[
\frac{n^2+R^2-2n}{4}\le\frac{n^2+n^{3/2}}{4}
\qquad\text{and}\qquad
\frac{n^2-R^2}{4}\ge\frac{n^2-n^{3/2}}{4}.
\]
Conditionally on $\sigma$, the number of equal-label edges is binomial with success probability $\frac12(1-\theta)$, and deleting these edges leaves a bipartite graph. Hence $\beta(G)\le\frac12(1-\theta)\frac{n^2+n^{3/2}}{4}+n^{3/2}\le\frac18n^2-\frac{\theta}{8}n^2+2n^{3/2}$. Likewise, the planted cut contains at least $\frac12(1+\theta)\frac{n^2-n^{3/2}}{4}-n^{3/2}\ge\frac18n^2+\frac\theta8n^2-2n^{3/2}$ edges. Since $\theta/8=4\eps$ and $n^{-1/2}+2e^{-4n}<0.01$ for $n\ge4\cdot10^4$, the lemma follows.
\end{proof}

\paragraph{Indistinguishability.}
We have shown that the two input distributions typically have different optimum values. We now show that a small fixed set of queries cannot reliably distinguish them. For this we compare the laws of the answers through their $\chi^2$-divergence, which controls the total variation distance and can be computed exactly for our two distributions.

Fix a set $F$ of distinct pairs, viewed as the edge set of a simple \emph{query graph} on $[n]$. Encode the answers by $X_e:=+1$ if $e\in E(G)$ and $X_e:=-1$ otherwise, and let $\cD_b^F$ be the law of $(X_e)_{e\in F}$ under $\cD_b$. A set $J\subseteq F$ is \emph{even} if every vertex has even degree in the graph $([n],J)$.

\begin{lemma}[$\chi^2$-divergence]\label{lem:chi2}
$1+\chi^2\bigl(\cD_1^F\,\Vert\,\cD_0^F\bigr)=\sum_{J\subseteq F\ \mathrm{even}}\theta^{2\abs{J}}$.
\end{lemma}

\begin{proof}
Under $\cD_0$ the answers are independent uniform signs, and under $\cD_1$, conditionally on $\sigma$, $\Prob(X_{uv}=x\mid\sigma)=\frac12(1-\theta x\sigma_u\sigma_v)$. Hence the likelihood ratio is $\Lambda(x)=\E_\sigma\prod_{uv\in F}(1-\theta x_{uv}\sigma_u\sigma_v)$, and $1+\chi^2(\cD_1^F\,\Vert\,\cD_0^F)=\E_0\Lambda(X)^2$. Let $\sigma'$ be an independent copy of $\sigma$. Integrating each independent uniform sign $X_{uv}$, we obtain
\[
\E_0\Lambda(X)^2=\E_{\sigma,\sigma'}\prod_{uv\in F}\bigl(1+\theta^2\sigma_u\sigma_v\sigma'_u\sigma'_v\bigr)=\E_\varrho\prod_{uv\in F}\bigl(1+\theta^2\varrho_u\varrho_v\bigr),
\]
where $\varrho_u:=\sigma_u\sigma'_u$ are independent uniform signs. Expanding the product, the term indexed by $J\subseteq F$ equals $\theta^{2\abs{J}}\prod_u\varrho_u^{\deg_J(u)}$, whose expectation is $\theta^{2\abs{J}}$ if $J$ is even and $0$ otherwise.
\end{proof}

The preceding identity reduces the divergence bound to counting even subgraphs of the query graph. We next bound this count in terms of the number of queried pairs.

\begin{lemma}[Counting even subgraphs]\label{lem:even-count}
A simple graph with $t\ge1$ edges has no nonempty even subgraph with fewer than three edges, and at most $(2\sqrt{2t})^k$ even subgraphs with exactly $k$ edges for every $k\ge3$.
\end{lemma}

\begin{proof}
Every nonempty even edge set is an edge-disjoint union of cycles (Veblen's theorem), and in a simple graph every cycle has length at least three. Let $B$ be the adjacency matrix, with eigenvalues $\lambda_i$, so that $\sum_i\lambda_i^2=\tr(B^2)=2t$. Each cycle of length $\ell\ge3$ corresponds to $2\ell$ closed walks of length $\ell$, so the number of such cycles is at most
\[
\frac{\tr(B^\ell)}{2\ell}\le\sum_i\abs{\lambda_i}^\ell\le\Bigl(\sum_i\lambda_i^2\Bigr)^{\ell/2}=(2t)^{\ell/2}.
\]
Fix one ordered decomposition $(C_1,\dots,C_s)$ into cycles for each even set with $k$ edges; this assignment is injective. The lengths $(\abs{C_1},\dots,\abs{C_s})$ form a composition of $k$, of which there are $2^{k-1}$, and for each composition there are at most $\prod_j(2t)^{\abs{C_j}/2}=(2t)^{k/2}$ sequences of cycles. Hence there are at most $2^{k-1}(2t)^{k/2}\le(2\sqrt{2t})^k$ even sets with $k$ edges.
\end{proof}

\begin{proof}[Proof of \cref{thm:lower}]
Suppose, for a contradiction, that $T\le\frac{1}{128}\theta^{-4}$, and let $F$ be any set of at most $T$ distinct pairs; repeated queries can be discarded. By \cref{lem:chi2,lem:even-count},
\[
\chi^2\bigl(\cD_1^F\,\Vert\,\cD_0^F\bigr)\le\sum_{k\ge3}\bigl(2\sqrt{2T}\,\theta^2\bigr)^k\le\sum_{k\ge3}4^{-k}=\frac1{48},
\]
and therefore $\dTV(\cD_0^F,\cD_1^F)\le\frac12\sqrt{\chi^2}<0.08$. A randomized nonadaptive algorithm uses random bits $\omega$ that are independent of the input; its query set $F=F(\omega)$ is a function of $\omega$, and its output is a function of $\omega$ and $(X_e)_{e\in F}$. Hence the total variation distance between the laws of its output under $\cD_0$ and under $\cD_1$ is at most $\E_\omega\,\dTV(\cD_0^{F(\omega)},\cD_1^{F(\omega)})<0.08$.

Now suppose that $\mathsf{Alg}$ estimates $\beta$, let $n\ge10\eps^{-2}$, and note that $n\ge4\cdot10^4$ because $\eps\le1/64$. Consider the event that the output exceeds $\tau:=\frac18n^2-2\eps n^2$. Under $\cD_0$, with probability at least $\frac23-0.01$ the output is accurate and $\beta(G)\ge\frac18n^2-3n^{3/2}$ (\cref{lem:separation}), in which case the output is at least $\frac18n^2-3n^{3/2}-\eps n^2>\tau$ because $3n^{3/2}<\eps n^2$. Under $\cD_1$, with probability at least $\frac23-0.01$ the output is at most $\frac18n^2-3\eps n^2+2n^{3/2}<\tau$. The probabilities of the event under $\cD_0$ and $\cD_1$ thus differ by at least $\frac13-0.02>0.08$, a contradiction. Hence $T>\frac1{128}\theta^{-4}=\frac{1}{128\cdot32^4}\,\eps^{-4}$. The argument for $\mc$ is the same, with the threshold $\frac18n^2+2\eps n^2$.
\end{proof}

\paragraph{Examined vertices.}
The lower bound on examined vertices holds for all algorithms, adaptive or not. Its proof is the canonical-tester argument of Goldreich and Trevisan~\cite{GT03}, which uses only that $\mc$ and $\beta$ are invariant under relabeling the vertices. The idea is to simulate an algorithm that examines $s$ vertices by a nonadaptive one that reads the subgraph induced by $s$ random vertices, and then to apply \cref{thm:lower} to the simulation.

\begin{corollary}[Examined vertices]\label{cor:vertices}
There are absolute constants $c,C>0$ such that for every $\eps\in(0,1/64]$ and every $n\ge C\eps^{-2}$, every algorithm, adaptive or not, that estimates $\beta(G)$ (respectively, $\mc(G)$) within $\pm\eps n^2$ with probability at least $2/3$ on every $n$-vertex graph examines at least $c\,\eps^{-2}$ vertices in the worst case.
\end{corollary}

\begin{proof}
Let $\mathsf{Alg}$ be such an algorithm, and let $s\le n$ be the maximum, over $n$-vertex inputs and outcomes of its random bits, of the number of distinct vertices in its queries. The simulation $\mathsf{Alg}'$ chooses a uniformly random sequence $(v_1,\dots,v_s)$ of distinct vertices, queries all pairs among them, and then runs $\mathsf{Alg}$, mapping the $j$th distinct vertex name used by $\mathsf{Alg}$ to $v_j$ and answering each query of $\mathsf{Alg}$ from the subgraph induced by $\{v_1,\dots,v_s\}$.

To verify that this simulation preserves the estimation guarantee, we compare it with running the original algorithm on a uniformly relabeled graph.
Let $\pi$ be a uniformly random permutation of $[n]$, independent of the random bits of $\mathsf{Alg}$, and run $\mathsf{Alg}$ on the relabeled graph $\pi(G)$. Condition on the random bits of $\mathsf{Alg}$. The $j$th distinct name $u_j$ is determined by the history so far, and hence by the vertices $\pi^{-1}(u_1),\dots,\pi^{-1}(u_{j-1})$ of $G$; so $\pi^{-1}(u_j)$ is uniformly distributed among the remaining $n-j+1$ vertices, as is $v_j$ in $\mathsf{Alg}'$. The run of $\mathsf{Alg}$ inside $\mathsf{Alg}'$ therefore has the same law as the run of $\mathsf{Alg}$ on $\pi(G)$. Since $\beta(\pi(G))=\beta(G)$, the algorithm $\mathsf{Alg}'$ estimates $\beta(G)$ within $\pm\eps n^2$ with probability at least $2/3$ on every graph. It is nonadaptive and makes $\binom s2$ queries, so \cref{thm:lower} gives $\binom s2\ge c'\eps^{-4}$, where $c'$ is the constant $c$ of that theorem. Hence $s\ge\sqrt{2c'}\,\eps^{-2}$.
\end{proof}

\Cref{thm:lower,cor:vertices} together prove \cref{thm:bip-intro}\ref{it:bip-lower}. For adaptive algorithms the query graph depends on the answers, and conditioning on it biases them, so \cref{lem:chi2} does not apply. Since $T$ queries examine at most $2T$ vertices, \cref{cor:vertices} shows that adaptive algorithms need $\Omega(\eps^{-2})$ queries, so the adaptive query complexity lies between $\Omega(\eps^{-2})$ and $O(\eps^{-4})$.

\paragraph{Max-$r$CSPs and local Hamiltonians.}
To extend the bound on examined vertices beyond graphs, we first pad graph constraints with dummy variables to reach every arity, and then write Boolean Max-$r$CSPs as diagonal Hamiltonians. In both encodings, a query to the new instance is answered by at most one query to the old one. As in \cref{sec:csp}, a variable of a Max-$r$CSP is examined if it belongs to a queried set.

\begin{corollary}[Examined variables]\label{cor:csp-vertices}
Fix $r\ge2$. For all sufficiently small $\eps>0$ and all $N\ge C_r\eps^{-2}$, every algorithm that estimates the optimum of every Boolean Max-$r$CSP on $N$ variables with payoffs in $[0,1]$ within $\pm\eps N^r$ with probability at least $2/3$ examines $\Omega_r(\eps^{-2})$ variables in the worst case.
\end{corollary}

\begin{proof}
For $r=2$ this is \cref{cor:vertices}. For $r\ge3$, take an $n$-vertex graph $G$ and add $r-2$ disjoint groups of $n$ dummy variables. For each pair $\{u,v\}\subseteq V(G)$ and each choice of one variable from each dummy group, assign the payoff $\one[uv\in E(G)]\one[x_u\ne x_v]$ to the resulting $r$-set; all other tables are zero. The dummy labels do not affect any payoff. There are $N=(r-1)n$ variables, and the optimum is exactly $n^{r-2}\mc(G)$.

An oracle query to any table can be answered with at most one query to the corresponding pair of $G$. Consequently, an estimator with error $\eps N^r$ gives, after division by $n^{r-2}$, a graph estimator with error $(r-1)^r\eps n^2$, while examining no more original graph vertices than the number of variables examined by the CSP algorithm. Apply \cref{cor:vertices} with accuracy $(r-1)^r\eps$. If $N$ is not divisible by $r-1$, use $n=\lfloor N/(r-1)\rfloor$ and add the remaining isolated variables; the same argument changes only constants depending on $r$.
\end{proof}

\begin{corollary}[Examined particles]\label{cor:quantum-particles}
Fix integers $r,d\ge2$. For all sufficiently small $\eps>0$ and all $N\ge C_r\eps^{-2}$, every algorithm that estimates the minimum product-state energy, or the ground-state energy, of every $r$-local Hamiltonian on $N$ particles of local dimension $d$, with at most one term of norm at most one per subset, within $\pm\eps N^r$ with probability at least $2/3$ examines $\Omega_r(\eps^{-2})$ particles in the worst case. A particle is examined if it belongs to a queried local interaction.
\end{corollary}

\begin{proof}
Let $g$ be a Boolean Max-$r$CSP with payoffs in $[0,1]$, and fix a surjection $\pi\colon[d]\to\{0,1\}$. For $e=\{i_1<\cdots<i_r\}$, define a diagonal local interaction by
\[
h_e:=-\sum_{a\in[d]^r}g_e\bigl(\pi(a_1),\ldots,\pi(a_r)\bigr)\lvert a\rangle\langle a\rvert,
\qquad H:=\sum_e h_e.
\]
Each $h_e$ has operator norm at most one. Since $H$ is diagonal in the computational basis, $E_0(H)$ is its smallest diagonal entry. The entry at a basis vector $\lvert b\rangle$ with $b\in[d]^N$ is the negative of the payoff of the Boolean assignment $(\pi(b_i))_i$, and every Boolean assignment arises in this way because $\pi$ is surjective. Thus $E_0(H)=-\Opt_g$. A minimizing basis vector is a product state, so $P(H)=E_0(H)=-\Opt_g$ as well. These identities also hold for every restriction. A query to $h_e$ is simulated by one query to $g_e$ and examines the same variables. Negating either energy estimate therefore gives a CSP optimum estimate with the same error and success probability. Apply \cref{cor:csp-vertices}.
\end{proof}

\section*{AI Disclosure}
Generative AI tools, especially GPT 5.6, GPT 6 and Opus 5.5, assisted with many parts of the mathematical development and exposition.
These tools were also used for writing the early-stage manuscript of this paper.
Despite this assistance, the authors put significant effort into verifying the results and making the text accessible to human readers, and retain full responsibility for the final text and results.

\clearpage
\bibliographystyle{alphaurl}
\bibliography{references}
\end{document}